\documentclass[11pt]{article}
\usepackage[margin=1in]{geometry}
\usepackage[colorlinks=true,
            linkcolor=black,
            citecolor=black,
            urlcolor=black]{hyperref}

\usepackage{setspace}

\usepackage{subcaption}

\usepackage{fixmath}
\usepackage{bm}
\usepackage{amsbsy}
\usepackage{color}
\usepackage{verbatim}
\usepackage{multirow}
\usepackage{amssymb}
\usepackage{amsthm}
\usepackage{amsmath}
\usepackage{array}
\usepackage{mathtools}

\allowdisplaybreaks

\usepackage{graphicx}
\usepackage{tikz}
\usetikzlibrary{positioning}
\usepackage{xcolor}

\usepackage{thm-restate}
\usepackage{algpseudocode}
\usepackage{algorithm}
\usepackage{algorithmicx}

\newtheorem{theorem}{Theorem}
\newtheorem{lemma}{Lemma}

\newtheorem{definition}{Definition}

\newtheorem{claim}{Claim}

\usepackage{natbib}
\usepackage{booktabs}
\usepackage{makecell}

\newcommand{\size}[1]{\ensuremath{|#1|}}
\newcommand{\ceil}[1]{\ensuremath{\lceil#1\rceil}}
\newcommand{\floor}[1]{\ensuremath{\lfloor#1\rfloor}}

\newcommand{\Ceil}[1]{\ensuremath{\left\lceil#1\right\rceil}}

\newcommand{\lra}[1]{\ensuremath{(#1)}}
\newcommand{\lrc}[1]{\ensuremath{\{#1\}}}
\newcommand{\lrd}[1]{\ensuremath{\langle #1\rangle}}

\newcommand{\lrA}[1]{\ensuremath{\left(#1\right)}}
\newcommand{\lrC}[1]{\ensuremath{\left\{#1\right\}}}

\def\B{\mathcal{B}}
\def\F{\mathcal{F}}
\def\HH{\mathcal{H}}
\def\O{\mathcal{O}}
\def\P{\mathcal{P}}
\def\R{\mathcal{R}}
\def\S{\mathcal{S}}
\def\T{\mathcal{T}}

\def\poly{\text{poly}}

\allowdisplaybreaks
\usepackage{multicol}
\newenvironment{claimproof}{\begin{proof}[Claim Proof]}{\end{proof}}
\newtheorem{property}{Property}
\def\capy{\mathrm{cap}}
\def\OPT{\mathrm{opt}}
\def\fOPT{\mathrm{flowopt}}
\def\Odd{\mathrm{Odd}}

\title{Improved Approximation Algorithms for the Multiple-Depot Split Delivery Vehicle Routing Problem}
\author{Jingyang Zhao$^{1}$, Yonghang Su$^{2}$, Mingyu Xiao$^{2}$\thanks{Corresponding author.} \\
$^{1}$Kyung Hee University \\
$^{2}$University of Electronic Science and Technology of China
}
\date{}
\begin{document}
\maketitle
\begin{abstract}
The multiple-depot split delivery vehicle routing problem is a challenging optimization problem with broad applications in logistics and transportation.
The goal is to serve customers' demand using a limited fleet of capacitated vehicles stationed at multiple depots, allowing each customer's demand to be split and served by multiple vehicles, while minimizing the total travel cost.
Parameterized by the number of depots, the previous best-known result was a slice-wise polynomial-time $6$-approximation algorithm (INFORMS J. Comput. 2023), and whether this ratio could be improved remained an open question.
We resolve this question by proposing a fixed-parameter tractable (FPT) $(2\alpha+1+\varepsilon)$-approximation algorithm for any constant $\varepsilon>0$, where $\alpha<3/2$ denotes the best-known ratio for the traveling salesman problem (TSP).
Our algorithm enumerates partitions of connected components formed by low-cost edges to construct a low-cost cycle cover, then extracts paths, and assigns them to vehicles through a minimum-cost flow method.
The cycle-cover technique also yields an FPT $(\alpha+\varepsilon)$-approximation for the multiple-depot TSP.
We further propose a simple parameterized $5$-approximation algorithm based on the structural properties of vehicle capacities, which achieves polynomial running time for a specific setting that appears in existing benchmark instances.
In addition, we develop a bi-factor approximation algorithm that balances minor vehicle capacity violations against reductions in travel cost or gains in computational efficiency.
Finally, our computational experiments demonstrate that the proposed methods exhibit complementary strengths across various instance types and achieve competitive solution quality.
\end{abstract}

\section{Introduction}
The (capacitated) vehicle routing problem~(VRP) is a classical and extensively studied routing problem~\citep{books/siam/TothV14}. 
In this problem, we are given a set of customers with integer demand and a depot that provides an \emph{unlimited} number of vehicles with the same integer capacity (while some experimental works consider a limited fleet, most approximation studies adopt this unlimited fleet setting).
The goal is to design vehicle routes to serve all customers such that the total travel cost is minimized. 
Each vehicle route must begin and end at the depot, and the vehicle must deliver at most its capacity of demand to customers.
There are two common variants of the VRP: \emph{splittable} and \emph{unsplittable}.
In the splittable VRP, each customer's demand can be served by multiple vehicles, whereas in the unsplittable VRP, each customer's demand must be served entirely by a single vehicle.
The practice of allowing split deliveries is practically well-justified; as shown by \citet{DBLP:journals/transci/ArchettiSS06}, relaxing the unsplittable constraint can yield up to a 50\% reduction in total routing costs.
When the model is extended to allow multiple depots, it becomes the multiple-depot VRP (MD-VRP), which has also received much attention in the literature~\citep{journals/cie/MontoyaTorresFIJH15}.

However, in real-world scenarios, the number of vehicles at each depot may be limited.
To address this situation, the split delivery VRP (SDVRP), a variant of the splittable VRP, was introduced~\citep{DBLP:journals/transci/DrorT89,journals/nrl/DrorT90}.
The SDVRP has since been studied extensively~\citep{journals/ior/BelenguerMM00,journals/transci/GamstLR24,journals/ijoc/LinHJMSL25,journals/or/KonstantakopoulosGK22}, and a comprehensive survey can be found in~\citep{DBLP:journals/eor/ArchettiBS14a}.
Analogously, when multiple depots are involved, the problem is known as the multiple-depot SDVRP (MD-SDVRP)~\citep{journals/transci/GouveiaLR23,journals/ijoc/LaiXXD23}.

In the MD-SDVRP~\citep{journals/ijoc/LaiXXD23}, we are given an undirected complete graph $G=(V, E)$ with vertex set $V=J\cup D$ and edge set $E$. 
The set $J$ denotes $n$ customers, and $D$ denotes $k$ depots.
Each customer $v\in J$ has a demand $q_v\in\mathbb{N}^+$, and each depot $u\in D$ owns $r_u\in\mathbb{N}^+$ vehicles with identical capacity $Q\in\mathbb{N}^+$.
Let $L$ denote the set of all vehicles, and let $m=\size{L}$.
Each edge $e\in E$ is associated with a non-negative cost $c(e)$, and the edge cost function $c$ is a metric, satisfying the triangle inequality. 
The goal is to design vehicle routes to serve all customers while minimizing the total travel cost.
Notably, when $k=1$, the MD-SDVRP becomes the SDVRP.

In this paper, we study parameterized approximation algorithms for the MD-SDVRP.
For a minimization problem $\Pi$, an algorithm is a $\rho$-approximation algorithm if, for any instance $I$ of $\Pi$, it computes a solution of cost at most $\rho$ times the optimal value in polynomial time $\poly(\size{I})$, where $\size{I}$ denotes the input size.
The quantity $\rho$ is referred to as the \emph{approximation ratio}.
With respect to some parameter $p\in\mathbb{N}^+$, an algorithm is a slice-wise polynomial-time~(XP) $\rho$-approximation algorithm if it runs in $\poly(\size{I})$ time for every fixed $p=\O(1)$ and achieves an approximation ratio of $\rho$; it is a fixed-parameter tractable~(FPT) $\rho$-approximation algorithm if it runs in $\O(f(p))\cdot \poly(\size{I})$ time for some computable function $f(p)$ and achieves an approximation ratio of $\rho$.

We also consider bi-factor approximation algorithms for the MD-SDVRP.
The notion of bi-factor approximation may vary across problems. 
For the MD-SDVRP, we define a bi-factor $(\rho,\sigma)$-approximation algorithm as a polynomial-time algorithm that computes a solution with cost at most $\rho$ times the optimal value (under the original capacity constraints), while allowing each vehicle to exceed its capacity $Q$, but with the total load on each vehicle bounded by $\sigma\cdot Q$, where $\sigma\geq 1$.
Analogously, we define XP and FPT bi-factor approximation algorithms for the MD-SDVRP.

\subsection{Related Work}
We review the existing approximation algorithms for the MD-SDVRP and related routing problems.

\paragraph{The VRP and the SDVRP.}
We first consider the VRP and the SDVRP. The VRP can be viewed as a special case of the SDVRP in which the depot is endowed with an unlimited number of vehicles.

For the uncapacitated case $Q=\infty$, by the triangle inequality, both the VRP and the SDVRP are equivalent to the traveling salesman problem~(TSP).
For the TSP, \citet{reports/cmu/Christofides76} and \citet{journals/us/Serdyukov78} independently proposed a $1.5$-approximation algorithm, and the ratio has been improved to less than $1.5-10^{-36}$ by~\citet{DBLP:conf/stoc/KarlinKG21,DBLP:conf/ipco/KarlinKG23}. Hereafter, we let $\alpha$ denote the current best-known approximation ratio for the TSP.

For the capacitated case $Q<\infty$, \citet{journals/moor/HaimovichK85} introduced the classical iterated tour partition~(ITP) algorithm for the splittable VRP, which is based on partitioning an $\alpha$-approximate TSP tour and achieves an approximation ratio of $\alpha+1$. 
\citet{journals/orl/AltinkemerG87} extended this approach to obtain a $(\alpha+2)$-approximation algorithm for the unsplittable VRP.
Later, \citet{journals/mp/BlauthTV23} improved these bounds to $\alpha+1-\varepsilon_\alpha$ for the splittable VRP and $\alpha+2-2\varepsilon_\alpha$ for the unsplittable VRP, where $\varepsilon_\alpha$ is a positive constant, greater than $\frac{1}{3000}$ when $\alpha=1.5$. 
For the unsplittable VRP, \citet{journals/moor/FriggstadMRS25} obtained a combinatorial $(\alpha+1.75)$-approximation algorithm and an $(\alpha+1+\ln2+\varepsilon)$-approximation algorithm for any constant $\varepsilon>0$ based on linear programming~(LP) rounding. Both ratios can be further improved by a small constant using the result in \citep{journals/mp/BlauthTV23}.
Recently, \citet{DBLP:journals/corr/abs-2601-21660} combined these two approaches and showed that the approximation ratio for the unsplittable VRP can be improved to $\alpha+1.6759$ for the case of general vehicle capacity and $\alpha+1.5897$ for the case of fixed vehicle capacity.

For the SDVRP, as noted by~\citet{journals/ijoc/LaiXXD23}, no constant-factor approximation was previously known. 
To the best of our knowledge, the only known result is a 4-approximation algorithm, which can be directly derived from their algorithm for the MD-SDVRP.

All the above results focus on general metrics. There is also a line of work on structured metrics~\citep{conf/iccmso/Chen23}, such as Euclidean spaces~\citep{journals/algorithmica/DasM15,journals/talg/JayaprakashS23,conf/soda/FriggstadGM26}, tree metrics~\citep{journals/talg/MathieuZ23,journals/mp/MathieuZ25}, and so on.

\paragraph{The MD-VRP and the MD-SDVRP.}
For the uncapacitated case $Q=\infty$, by the triangle inequality, both the MD-VRP and the MD-SDVRP are equivalent to the multiple-depot TSP (MD-TSP).
For the MD-TSP, \citet{journals/tase/RathinamSD07} proposed a 2-approximation algorithm, and \citet{journals/orl/XuXR11} improved the ratio to $2-1/k$ for $k\geq 2$. 
Moreover, parameterized by $k$, \citet{journals/ijoc/XuR15} obtained an XP $1.5$-approximation algorithm based on edge exchanges, with running time $\O(\size{V}^{k+2}\size{E}^{k-1})$. 
Later, \citet{DBLP:journals/siamcomp/TraubVZ22} obtained an XP $(\alpha+\varepsilon)$-approximation algorithm with running time $\size{V}^{\O(k/\varepsilon)}$ for any constant $\varepsilon>0$. 
Their algorithm yields the same approximation ratio for a broad class of routing problems, specifically the $\Phi$-TSP with bounded \emph{interface} size.
Recently, \citet{conf/esa/DeppertKM23} obtained a randomized FPT $(1.5+\varepsilon)$-approximation algorithm with running time $(1/\varepsilon)^{\O(k\log k)}\size{V}^{\O(1)}$, based on a reduction to the rural postman problem~(RPP)~\citep{DBLP:journals/jcss/GutinWY17}.
Furthermore, for a generalized MD-TSP, where only $s$ out of $k$ depots can be selected to serve customers, \citet{DBLP:journals/eor/XuR17} extended the algorithm of \citet{journals/orl/XuXR11} to achieve a $(2-1/(2s))$-approximation algorithm.

For the capacitated case $Q<\infty$, \citet{journals/ijoc/LiS90} proposed a $(2\alpha+1)$-approximation algorithm for the splittable MD-VRP and a $(2\alpha+2)$-approximation algorithm for the unsplittable MD-VRP. 
Both methods partition an $\alpha$-approximate TSP tour within a modified graph where all depots are contracted into a single super-depot. 
Utilizing a forest-partitioning technique, \citet{journals/transci/HarksKM13} improved this ratio to $4$ for the unsplittable MD-VRP.
Later, based on extending the results of VRP, \citet{DBLP:journals/tcs/ZhaoX25a} improved the ratio to $4-\frac{1}{1500}$ for the splittable MD-VRP and to $4-\frac{1}{50000}$ for the unsplittable MD-VRP. 
Very recently, based on rounding a new LP relaxation, \citet{DBLP:journals/corr/abs-2510-05321} further improved the ratio to $3.9365$ for the \emph{unit-demand} MD-VRP, a special case of the unsplittable MD-VRP in which each customer has unit demand.
Moreover, parameterized by $Q$, \citet{DBLP:journals/tcs/ZhaoX25a} obtained an XP $(3+\ln2-\max\{\Theta(\frac{1}{\sqrt{Q}}),\frac{1}{9000}\})$-approximation algorithm for the splittable MD-VRP and an XP $(3+\ln2-\Theta(\frac{1}{\sqrt{Q}}))$-approximation algorithm for the unsplittable MD-VRP, both of which run in $\size{V}^{\O(Q)}$ time.

For the MD-SDVRP, \citet{phd/sfu/Hu09} obtained an exact algorithm on path metrics, and a $2$-approximation algorithm on tree metrics. 
The latter result also yields an $\O(\log\size{V})$-approximation algorithm on general metrics via \emph{tree embedding} techniques~\citep{DBLP:journals/jcss/FakcharoenpholRT04}. 
Recently, \citet{journals/ijoc/LaiXXD23} showed that solutions obtained via approximation algorithms for the MD-VRP may require an arbitrarily larger number of vehicles than the minimum number of vehicles required. To address this, they proposed a \emph{pseudo}-XP $(6-4/k)$-approximation algorithm parameterized by $k\geq 2$, running in time $\O(\size{V}^{k-1}\size{E}^{2(k-1)}\max\{\size{V}^3,\size{L}^3\})$. 
Their algorithm is pseudo-XP as the value $\size{L}$ may be super-polynomial in $\size{V}$.
Their analysis focuses on the case $k\geq 2$. 
When $k=1$, their algorithm achieves an approximation ratio of $4$ with running time $\O(\max\{\size{V}^3,\size{L}^3\})$.
Whether the approximation ratios can be improved remains open.

The capacitated location routing problem~(CLRP) is closely related to the MD-VRP, which integrates elements of facility location problems~(FLPs) and MD-VRPs~\citep{journals/anor/SchneiderD17,journals/ijoc/ContardoCG14,journals/ijoc/HeHW25}. 
In the CLRP, each depot $u\in D$ is additionally associated with a non-negative opening cost, and the goal is first to select a set of depots to open and then to route the unlimited number of vehicles from the opened depots to serve customers, while minimizing the total cost, including both the cost of opening depots and the travel cost of serving customers. When the opening cost of each depot is zero, the CLRP becomes the MD-VRP, since one can open all depots at no cost. 
For the unsplittable CLRP, \citet{journals/transci/HarksKM13} proposed a $4.38$-approximation algorithm. 
Note that this algorithm is the one achieving an approximation ratio of $4$ for the unsplittable MD-VRP. They further extended the approach to three variants of the CLRP: prize-collecting, grouping, and cross-docking.
Later, \citet{DBLP:conf/ijcai/00010W24} improved the ratio to $4.169$ for the unsplittable CLRP and to $4.091$ for the splittable CLRP.

\citet{journals/eor/HeineDM23} studied a variant of the splittable CLRP, where each depot $u\in D$ is additionally assigned a capacity $f(u)\geq 0$, meaning that the total customer demand served by vehicles from $u$ is at most $f(u)$.
When the opening cost of each depot is zero, the problem becomes the capacitated MD-VRP (CMD-VRP)---a problem that, as we establish later in this work, shares a profound connection with the MD-SDVRP.
They proposed a bi-factor $(4+2\delta/\varepsilon, 1+\varepsilon)$-approximation algorithm for any constant $\varepsilon>0$, where $\delta$ denotes the approximation ratio for the capacitated FLP (CFLP). 
Their algorithm computes a solution with cost at most $4+2\delta/\varepsilon$ times the optimal value (under the original depot capacity constraints), while allowing the capacity constraint of each depot to be violated by at most $\varepsilon \cdot Q$ (rather than $\varepsilon \cdot f(u)$).
Currently, $\delta = 5$ for the general case~\citep{conf/esa/BansalGG12}, $\delta = 3$ for the uniform depot capacities case~\citep{journals/mp/AggarwalLBGGGJ13}, and $\delta=1$ when the opening cost of each depot is zero.
Hence, the CMD-VRP admits a bi-factor $(4+2/\varepsilon,1+\varepsilon)$-approximation algorithm.

Another problem related to the unsplittable CLRP, introduced by~\citet{journals/networks/GortzN16}, considers locating $k$ depots at no cost and then routing vehicles from these depots to serve all customers while minimizing the total travel cost. 
They proposed a $(12+\varepsilon)$-approximation algorithm for any constant $\varepsilon > 0$ based on local search techniques.

The capacitated rooted tree cover problem~(CRTCP)~\citep{DBLP:conf/approx/DasJK20} is also closely related to the MD-SDVRP.
The goal is to compute $k$ trees rooted at the given depots that cover all customers, such that each tree covers at most $Q$ customers and the total edge cost is minimized.
By doubling the tree edges and shortcutting, any $\O(1)$-approximation algorithm for the CRTCP yields an $\O(1)$-approximation algorithm for the MD-SDVRP when each customer has unit demand and each depot owns a single vehicle.
However, $\O(1)$-approximation algorithms are known only for the \emph{makespan} version of the CRTCP~\citep{DBLP:conf/approx/DasJK20}, i.e., minimizing the maximum cost among all trees.

Finally, a problem more general than the CRTCP, introduced by~\citet{DBLP:conf/stacs/AdamaszekAM16}, is the airport and railway problem~(ARP).
In this problem, each vertex has an opening cost for being selected as a depot.
The objective is to first select a set of vertices to open as depots and then find a set of trees covering all vertices, such that each tree contains at least one opened depot. Moreover, each tree must cover at most $Q$ vertices, and the goal is to minimize the sum of the depot opening costs and the total edge cost of all trees.
The CRTCP is a special case of the ARP, in which exactly $k$ vertices have zero opening cost and all others have infinite opening cost. 
For the ARP, most existing approximation algorithms focus on structured metrics~\citep{DBLP:conf/stacs/AdamaszekA0M18,DBLP:journals/ipl/JowhariN25}, and the best-known result on general metrics is an $\O(\log\size{V})$-approximation algorithm based on tree embedding techniques~\citep{DBLP:journals/jco/SalavatipourT25}. 

\subsection{Our Results}

Our first contribution is an FPT approximation algorithm, denoted as \textsc{Alg.1}, for the MD-SDVRP parameterized by the number of depots $k$.
By taking any polynomial-time $\alpha$-approximation algorithm for the TSP as a black box, for any constant $\varepsilon>0$, \textsc{Alg.1} achieves a ratio of $2\alpha+1+\varepsilon$ with running time $2^{\O(\frac{k\log k}{\varepsilon})}\cdot \O(|V|f(|V|)+\size{V}^4\log\size{V})$, where $f(|V|)$ denotes the running time of the TSP algorithm.
Using the result of \citet{DBLP:conf/stoc/KarlinKG21,DBLP:conf/ipco/KarlinKG23}, we obtain an FPT $(4-10^{-36})$-approximation, improving upon the previous $6$-approximation algorithm of \citet{journals/ijoc/LaiXXD23} with running time $\O(\size{V}^{k-1}\size{E}^{2(k-1)}\max\{\size{V}^3,\size{L}^3\})$.
Note that our approximation ratio almost matches the classical $(2\alpha+1)$-approximation ratio for the splittable MD-VRP~\citep{journals/ijoc/LiS90}. 
Therefore, to further significantly improve our ratio, one may need to first investigate how to design significantly improved (FPT) approximation algorithms for the splittable MD-VRP.

Our framework also yields an $(\alpha+\varepsilon)$-approximation algorithm for the MD-TSP with running time $2^{\O(\frac{k}{\varepsilon})}\cdot\O(f(|V|)+|V|^2\log|V|)$, improving the randomized FPT $(3/2+\varepsilon)$-approximation~\citep{conf/esa/DeppertKM23} and the running time of the XP $(\alpha+\varepsilon)$-approximation algorithm~\citep{DBLP:journals/siamcomp/TraubVZ22}.

Our second contribution is a $5$-approximation algorithm, denoted as \textsc{Alg.2}, for the MD-SDVRP parameterized by $Q$ and $k$.
Its running time is $\O(\min\{Q^{k-1}, \binom{mQ-\sum_{v\in J}q_v+k-1}{k-1}\} \cdot\size{V}^4\log\size{V})$, which is FPT with a single-exponential dependence on $k$ when $Q=\O(1)$, and is polynomial time when $mQ-\sum_{v\in J}q_v=\O(1)$.
This efficiency is particularly relevant for practical benchmarks.
For example, among the 63 instances of the SD Set~\citep{journals/ijoc/LaiXXD23}, 60 satisfy the zero-excess capacity condition $mQ-\sum_{v\in J}q_v=0$.
On these instances, \textsc{Alg.2} yields a $5$-approximation in $\O(\size{V}^4\log\size{V})$ time.

Our third contribution is a bi-factor approximation algorithm for the MD-SDVRP, denoted as \textsc{Alg.3}.
Our \textsc{Alg.3} relates the MD-SDVRP to the CMD-VRP.
We obtain a $(6+2/\varepsilon+2\varepsilon/(1+\varepsilon),1+\varepsilon)$-approximation by using the $(4+2/\varepsilon,1+\varepsilon)$-approximation for the CMD-VRP~\citep{journals/eor/HeineDM23}.
We also obtain an $(\alpha+1-\varepsilon_\alpha)$-approximation
for the SDVRP, matching the bound for the splittable
VRP~\citep{journals/mp/BlauthTV23}.

Our final contribution is an experimental evaluation of the algorithms.
For implementation, we propose a practical variant of \textsc{Alg.1}, denoted as \textsc{Alg.1+}, which achieves an approximation ratio of $9$ with running time $2^{\O(k\log k)}\cdot\O(\size{V}^4\log\size{V})$.
We compare \textsc{Alg.1+} and \textsc{Alg.2} with two algorithms from \citet{journals/ijoc/LaiXXD23}.
The results show that the proposed methods exhibit complementary strengths across different classes of instances and achieve competitive solution quality.

\section{Preliminaries}\label{sec2}
In the MD-SDVRP, the input instance is denoted by $I=(G,c,q,r,Q)$, where $G=(V, E)$ is an undirected complete graph with vertex set $V=J\cup D$ and edge set $E$. 
The set $J=\{v_1,\dots,v_n\}$ denotes the customers, and $D=\{u_1,\dots,u_k\}$ denotes the depots. 
There is a cost function $c: V\times V\to \mathbb{R}_{\geq0}$, which is a \emph{metric} satisfying $c(x,x)=0$, $c(x,y)=c(y,x)$, and $c(x,y)\leq c(x,z)+c(z,y)$ (the triangle inequality) for all $x,y,z\in V$.
There is a demand function $q:J\to\mathbb{N}^+$, where $q_v$ is the demand of customer $v$. 
Each depot $u$ owns $r_u\in\mathbb{N}^+$ vehicles with identical capacity $Q\in\mathbb{N}^+$, and each vehicle serves at most $Q$ demand.
Let $L=\{\tau_1,\dots,\tau_m\}$ be the set of all vehicles, where $m=\sum_{u\in D}r_u$, and let $\varphi(\tau) \in D$ denote the depot where vehicle $\tau$ is located.
We assume $\sum_{v\in J}q_v\leq mQ$ for feasibility.

A \emph{walk} $W$ in $G$, denoted by $v_1v_2\dots v_\ell$, is a sequence of vertices in $V$, where a vertex may be repeated, and each consecutive pair of vertices is connected by an edge.
Let $E(W)$ denote the (multi-)set of edges traversed by $W$. Then, the cost of $W$ is defined as $c(W)=\sum_{e\in E(W)}c(e)$.

A \emph{path} is a walk where no vertex appears more than once. 
A \emph{closed} walk is a walk where the first and the last vertices are the same, and
a \emph{cycle} is a walk where only the first and the last vertices are the same.
A cycle (or path) $C$ containing a single vertex is called \emph{trivial}, for which we define $E(C)=\emptyset$.
A perfect \emph{matching} $M$ in $G$ is a set of vertex-disjoint edges that spans all vertices in $V$.
A \emph{cycle cover} $\B$ in $G$ is a set of vertex-disjoint cycles that spans all vertices in $V$. A TSP tour in $G$ is a cycle that spans all vertices in $V$.

Given a closed walk, by the triangle inequality, one can skip repeated vertices along the walk to obtain a cycle of non-increasing cost.
Such an operation is called \emph{shortcutting}.
We define a \emph{route} or \emph{tour} for a vehicle from depot $u$ to be a cycle in the form of $u v_1v_2\dots v_\ell u$, where each $v_i$ is a customer. 
A tour containing a single customer, e.g., $uv_1u$, is called an \emph{atomic} tour.

For any (multi-)graph $H=(V_H,E_H)$ and any vertex subset $V_s\subseteq V_H$, let $H[V_s]$ denote the subgraph of $H$ induced by $V_s$. 
In $H$, the \emph{degree} of a vertex $v\in V_H$ is the number of edges in $E_H$ incident to it. 
Let $\Odd(H)$ denote the set of odd-degree vertices in $V_H$.
The graph $H$ is called an \emph{even-degree graph} if $\Odd(H)=\emptyset$.
Moreover, $H$ is called an \emph{Eulerian graph} if it is connected in addition to being even-degree.
Given an Eulerian graph $H$, an \emph{Eulerian tour}, i.e., a closed walk that traverses each edge in $E_H$ exactly once, can be computed in $\O(\size{V_H}+\size{E_H})$ time~\citep{books/mit/CormenLRS22}. 

For any (multi-)graph $H$ whose vertices belong to $V$, we use $V(H)$, $J(H)$, $D(H)$, $E(H)$, and $c(H)$ to denote the vertex set, customer set, depot set, edge set, and total edge cost of $H$, respectively.
This notation is also used for any walk or tree, as well as for any set of walks, trees, or components.
Note that all components considered in this paper are connected.

For any (multi-)set of edges $E'$ with elements in $E$, let $G_{E'}=(V,E')$ denote the corresponding (multi-)graph induced by $E'$, which can be viewed as a set of components, denoted by $\S_{E'}$. (Recall that a component is a maximal connected subgraph.)
For a component $C$, we may view it as a graph with vertex set $V(C)$ and edge set $E(C)$.

We also consider a \emph{network} $N=(V_N,E_N)$, where each edge is directed and associated with a capacity. 
We let $\lrd{x,y}_z$ denote a directed edge from $x$ to $y$ with capacity $\capy\lrd{x,y}=z\in\mathbb{N}^+$.

The MD-SDVRP is defined as follows.
\begin{definition}[The MD-SDVRP]
Given $I=(G,c,q,r,Q)$, the goal is to construct a set of tours $\T=\{T_\tau\}_{\tau\in L}$ with a demand assignment $\lambda: J\times\T\to \mathbb{N}$ such that
\begin{multicols}{2}
\begin{enumerate}
\item $T_\tau$ is a tour starting at $\varphi(\tau)$ for each $\tau\in L$,
\item $\sum_{v\in J(T)}\lambda_{v,T}\leq Q$ for each $T\in \T$,
\item $\lambda_{v,T}=0$ for each $T\in \T$ and $v\in J\setminus J(T)$,
\item $\sum_{T\in\T}\lambda_{v,T}=q_v$ for each $v\in J$,
\end{enumerate}
\end{multicols}
and $c(\T)=\sum_{T\in\T}c(T)$ is minimized. 
\end{definition}
In the definition, each customer may be served by multiple tours.
By the triangle inequality, we may assume without loss of generality that $\lambda_{v,T} > 0$ for all $v \in J(T)$.
When $J(T_\tau)=\emptyset$, the vehicle $\tau$ delivers no goods to customers. In this case, we say that $\tau$ is \emph{inactive} in $\T$; otherwise, it is \emph{active}.

Since $Q$ and $|L|$ may be super-polynomial in $|V|$, we allow solutions to be represented compactly by recording identical tours with the same demand assignment together with their multiplicities. Inactive vehicles are represented implicitly.

Fix an optimal solution $\T^*=\{T^*_\tau\}_{\tau\in L}$ to the problem, and let $\OPT=c(\T^*)$ denote its cost.
The solution can be viewed as a set of components $\S^*$.
In particular, the graph $G_{E(\T^*)}$ induced by the edges in $E(\T^*)$ can be uniquely decomposed into a set of components, which we denote by $\S^*$.

For any component $C$, define $\ell(C)=\ceil{\frac{\sum_{v\in J(C)}q_v}{Q}}$, which is the minimum number of vehicles needed to serve all customers in $J(C)$. For any set of components $\S$, define $\ell(\S)=\sum_{C\in\S}\ell(C)$.
Then, at least $\ell(\S^*)$ active vehicles exist in the optimal solution $\T^*$, and hence $\ell(\S^*)\leq m$.

With respect to $\S^*$, we call a set of components $\S$ \emph{good} if it satisfies the following condition: for any $C\in \S^*$, all vertices in $J(C)$ are contained in a single component in $\S$.
Thus, for any $C_i\in\S$, there exists a non-empty subset $\S^*_i\subseteq\S^*$ with $J(C_i) = \bigcup_{C\in\S^*_i}J(C)$.
Then, $\ell(\S)\leq \ell(\S^*)\leq m$.

\begin{property}\label{good}
Given a set of components $\S$, if one uses $\ell(C)$ vehicles to serve all customers in $J(C)$ for each $C\in\S$, the total number of vehicles used is at most $m$ when $\S$ is good.
\end{property}

The remainder of this paper is organized as follows.
Section~\ref{sec3} presents the FPT $(2\alpha+1+\varepsilon)$-approximation algorithm \textsc{Alg.1} and its application to the MD-TSP.
Section~\ref{sec5} presents the parameterized $5$-approximation algorithm \textsc{Alg.2}.
Section~\ref{sec6} presents the bi-factor approximation algorithm \textsc{Alg.3}.
Section~\ref{sec8} reports the experimental results.

\section{FPT Approximation Algorithms for the MD-SDVRP}\label{sec3}
We first give our FPT $(2\alpha+1+\varepsilon)$-approximation algorithm (\textsc{Alg.1}) for the MD-SDVRP parameterized by $k$.
Then, for the case $Q=\infty$ (i.e., the MD-TSP), we show that our method yields an FPT $(\alpha+\varepsilon)$-approximation algorithm.

\subsection{The Algorithm}\label{SC3.1}
\textsc{Alg.1} applies a given $\alpha$-approximation algorithm for the TSP as a black-box.
We assume that such an algorithm is available and runs in $f(\size{V})$ time for some function $f$.

At a high level, \textsc{Alg.1} first computes a good set of components $\S$ such that $\S$ is also a cycle cover in $G$.
Then, it constructs a solution $\T$ to the MD-SDVRP by using $\ell(C)$ vehicles to serve all customers in $J(C)$ for each $C\in\S$, which uses at most $m$ vehicles in total by Property~\ref{good}.
Since $\T^*$ is unknown, we cannot check whether a set of components is good or not.
\textsc{Alg.1} tries $2^{\O(\frac{k\log k}{\varepsilon})}$ possible sets of components, among which Lemma~\ref{lem:weighted-cover} shows that there exists one good set with small cost.
Hence, \textsc{Alg.1} computes a solution for each set and returns the best-found solution.

Next, we introduce the construction of component sets.

\subsubsection{The Component Sets}
For any constant $\varepsilon>0$, let $K\coloneqq k+\ceil{\frac{k}{\varepsilon}}$.
For any $\tau\geq0$, let $G_\tau=(V,E_\tau)$ denote the graph on $V$ containing all edges in $E$ whose cost is at most $\tau$.

Given the MD-SDVRP instance $I$, \textsc{Alg.1} first finds the smallest $\tau\in\lrc{0}\cup\lrc{c(e)}_{e\in E}$ such that $G_\tau$ has at most $K$ components, denoted by $\B$.
Then, it enumerates every possible partition $\mathfrak{B}$ of $\B$ such that $\size{\mathfrak{B}}\leq k$, $V(\B_i)\cap D\neq \emptyset$ for each $\B_i\in\mathfrak{B}$, and $\sum_{\B_i\in\mathfrak{B}}\ceil{\frac{\sum_{v\in J(\B_i)}q_v}{Q}}\leq m$.
There are at most $k^K = 2^{\O(\frac{k\log k}{\varepsilon})}$ partitions, which incurs $2^{\O(\frac{k\log k}{\varepsilon})}$ iterations.
For each $\mathfrak{B}$, it is transformed into a set of components $\S=\{C_i\}_{\B_i\in\mathfrak{B}}$, where $C_i$ is an $\alpha$-approximate TSP solution in $G[V(\B_i)]$.

To further transform $\S$ into a solution $\T$ to $I$, \textsc{Alg.1} calls the sub-algorithm \textsc{Transform}.
Note that \textsc{Transform} will use $\ell(\S)=\sum_{\B_i\in\mathfrak{B}}\ceil{\frac{\sum_{v\in J(\B_i)}q_v}{Q}}\leq m$ vehicles. 

\textsc{Alg.1} is described in Algorithm~\ref{alg1}.
We next show how to use $\S$ to construct a solution $\T$ to $I$.

\begin{algorithm}[!t]
\small
\caption{\textsc{Alg.1}: An FPT $(2\alpha+1+\O(\varepsilon))$-approximation algorithm for the MD-SDVRP}
\label{alg1}
\textbf{Input:} An instance $I=(G,c,q,r,Q)$, a constant $\varepsilon>0$, and an $\alpha$-approximation algorithm for the metric TSP. \\
\textbf{Output:} A feasible solution $\T_f$.
\begin{algorithmic}[1]

\State Set $K\coloneq k+\ceil{\frac{k}{\varepsilon}}$, and initialize $\mathfrak{T}\coloneq\emptyset$.

\State Find the smallest $\tau\in\lrc{0}\cup\lrc{c(e)}_{e\in E}$ such that $G_\tau$ has at most $K$ components, and let $\B$ be its component set.\label{alg:main-threshold}

\For{each partition $\mathfrak{B}$ of $\B$ with $\size{\mathfrak{B}}\leq k$, $V(\B_i)\cap D\neq \emptyset$ for each $\B_i\in\mathfrak{B}$, and $\sum_{\B_i\in\mathfrak{B}}\ceil{\frac{\sum_{v\in J(\B_i)}q_v}{Q}}\leq m$}\label{alg:main-partition}

\State For each $\B_i\in\mathfrak{B}$, apply the given $\alpha$-approximation algorithm for the TSP to $G[V(\B_i)]$ to obtain a cycle $C_i$.\label{alg:main-cycles}

\State Let $\S\coloneq\{C_i\}_{\B_i\in\mathfrak{B}}$, call \textsc{Transform} on $\S$ to obtain $\T$, and update $\mathfrak{T}\coloneq\mathfrak{T}\cup\{\T\}$.\label{alg:main-transform}

\EndFor

\State \Return $\T_f=\arg\min_{\T\in\mathfrak{T}}c(\T)$.
\end{algorithmic}
\end{algorithm}

\subsubsection{The Solution Constructions}
Since $\S$ is a cycle cover in $G$, it can be updated into a cycle cover in $G[J]$ by shortcutting. 
We now describe \textsc{Transform}.

\paragraph{Extracting the paths.}
\textsc{Alg.1} uses $\ell(C)$ vehicles to serve all customers in $J(C)$ for each $C\in\S$.
As shown in~\citep{journals/ijoc/LaiXXD23}, one possible approach is to \emph{greedily} extract $\ell(C)$ paths from $C$, where each of the first $\ell(C)-1$ paths is assigned exactly $Q$ demand and the last path is assigned $\sum_{v\in J(C)}q_v-(\ell(C)-1)\cdot Q$ demand, followed by satisfying the assigned demand of each path using one vehicle.
The details of the path extraction from $C$ are as follows.

Suppose $C=v_1\dots v_jv_1$. Define $q'_{v_i}\coloneq q_{v_i}$ for each $i$, and let $q'_C\coloneq\sum_{i=1}^jq'_{v_i}$. Initialize $p=i_0=1$.
Repeat the following procedure until $q'_C = 0$:
\begin{enumerate}
    \item find the minimum index $i_p$ such that $i_{p-1}\leq i_{p}\leq j$ and $\sum_{i=i_{p-1}}^{i_p}q'_{v_i}\geq M$, where $M=\min\{Q,q'_C\}$;
    \item extract a path $P=v_{i_{p-1}}\dots v_{i_p}$, and set $\lambda_{v_i,P}=q'_{v_i}$ for $i_{p-1}\leq i<i_p$, and $\lambda_{v_{i_p},P}=M-\sum_{i=i_{p-1}}^{i_p-1}q'_{v_i}$;
    \item update $q'_{v_i}\coloneq0$ for each $i_{p-1} \leq i<i_p$, $q'_{v_{i_p}}\coloneq q'_{v_{i_p}}-\lambda_{v_{i_p},P}$, $q'_C\coloneq q'_C-M$, and $p\coloneq p+1$.
\end{enumerate}

The above algorithm produces a set of paths $\P_C$ with $\size{\P_C}=\ell(C)$ and a demand assignment $\lambda$ such that the first $\ell(C)-1$ paths are each assigned exactly $Q$ demand and the last is assigned at most $Q$ demand. 
All customers in $J(C)$ can be served by satisfying the demand of each path in $\P_C$.

Hence, there are $\ell(\S)$ paths in $\bigcup_{C\in\S}\P_C$. 
By satisfying the demand of every path, one would obtain a feasible solution to the MD-SDVRP using exactly $\ell(\S)$ vehicles.

Note that each demand $q_v$ may be super-polynomial in $\size{V}$; thus, the above method may run in pseudo-polynomial time.
To address this, we adopt a slightly modified approach in \textsc{Alg.1}.

Consider a cycle $C\in\S$.
For each $v\in J(C)$ with $q_v\geq Q$, we extract a trivial path $v$ with \emph{load} $\floor{\frac{q_v}{Q}}$, which means that the demand of this trivial path needs to be satisfied using $\floor{\frac{q_v}{Q}}$ vehicles, i.e., it corresponds to $\floor{\frac{q_v}{Q}}$ trivial paths, each with demand $Q$. 
Hence, let $q'_v\coloneq q_v-\floor{\frac{q_v}{Q}}\cdot Q$.
Using the residual demand $q'_v$, we then apply the previous path extraction method to obtain a set of paths $\P'_C$ with $\size{\P'_C}=\ceil{\frac{\sum_{v\in J(C)}q'_v}{Q}}$ together with the corresponding assignment $\lambda'_{v,P}$ for each $P\in\P'_C$ and $v\in J(P)$.
Let the set of trivial paths be $J_C=\{v\in J(C)\mid q_v\geq Q\}$.
Then, all customers in $J(C)$ can be served by satisfying all paths in $J_C\cup\P'_C$.
Note that the total number of vehicles required to satisfy all paths in $J_C\cup\P'_C$ is $\sum_{v\in J_C}\floor{\frac{q_v}{Q}}+\size{\P'_C}$, which equals $\ell(C)$. 

\paragraph{Assigning the paths to vehicles.}
Let $J_\S=\bigcup_{C\in\S} J_C$ and $\P'_\S=\bigcup_{C\in\S} \P'_C$. To obtain a solution, we match each path in $\P'_\S$ with one vehicle and each path $v\in J_\S$ with $\floor{\frac{q_v}{Q}}$ vehicles.
The assignment is a minimum-cost bipartite $b$-matching problem:
each path $P\in\P'_{\S}$ requires one vehicle, each trivial path
$v\in J_{\S}$ requires $\floor{\frac{q_v}{Q}}$ vehicles, and each depot
$u\in D$ owns at most $r_u$ vehicles.
We solve this problem via its standard reduction to the minimum-cost maximum flow problem~(MCMFP).
First, we construct a network $N=(V_N,E_N)$, where we initialize $V_N\coloneq\{s,t\}\cup D$ and $E_N\coloneq\{\lrd{u,t}_{r_u}\mid u\in D\}$.
Then, for each trivial path $v\in J_\S$, we update $V_N\coloneq V_N\cup\{v\}$ and $E_N\coloneq E_N\cup\{\lrd{s,v}_{\floor{\frac{q_v}{Q}}}\}$;
for each path $P\in \P'_\S$, we update $V_N\coloneq V_N\cup\{P\}$ and $E_N\coloneq E_N\cup\{\lrd{s,P}_{1}\}$. 
Finally, for each $v\in J_\S$, $P\in \P'_\S$, and $u\in D$, we update $E_N\coloneq E_N\cup\{\lrd{v,u}_{\infty},\lrd{P,u}_{\infty}\}$.
Moreover, we define $c\lrd{v,u}=c(v,u)$ and $c\lrd{P,u}=\min_{w\in J(P)}c(w,u)$; for all other edges, we define the cost to be zero.
The construction of the network $N$ is illustrated in Figure~\ref{fig0}.

\begin{figure}[t]
\centering
\caption{An illustration of the network construction. The costs of the
colored edges are $c\lrd{s,v}=c\lrd{s,P}=c\lrd{u,t}=0$,
$c\lrd{v,u}=c(v,u)$, and
$c\lrd{P,u}=\min_{w\in J(P)}c(w,u)$.}
\label{fig0}
\begin{tikzpicture}[
    dot/.style={circle, fill, inner sep=1.5pt},
    edge/.style={-stealth, thick},
    every label/.style={inner sep=1pt},
    capacity label/.style={font=\scriptsize, fill=none, inner sep=1pt}
]
\node[dot, label=left:$s$] (S) at (0,0) {};
\node[dot, label=above:{$v\in J_{\S}$}] (v) at (2.8,0.42) {};
\node[dot, label=below:{$P\in\P'_{\S}$}] (P) at (2.8,-0.42) {};

\node[dot] (u_top) at (6.4,0.95) {};
\node[dot, label=above right:$u$] (u) at (6.4,0) {};
\node[dot] (u_bot) at (6.4,-0.95) {};
\node[font=\scriptsize, inner sep=0pt] at (6.4,0.39) {$\vdots$};
\node[font=\scriptsize, inner sep=0pt] at (6.4,-0.39) {$\vdots$};

\node[dot, label=right:$t$] (t) at (9.2,0) {};

\draw[edge] (v) -- (u_top);
\draw[edge] (v) -- (u_bot);
\draw[edge] (P) -- (u_top);
\draw[edge] (P) -- (u_bot);
\draw[edge] (u_top) -- (t);
\draw[edge] (u_bot) -- (t);

\draw[edge, red] (S) --
    node[capacity label, above, pos=0.45]
    {$\lrd{s,v}_{\floor{\frac{q_v}{Q}}}$} (v);
\draw[edge, red] (S) --
    node[capacity label, below, pos=0.45]
    {$\lrd{s,P}_1$} (P);

\draw[edge, blue] (v) --
    node[capacity label, above, pos=0.36]
    {$\lrd{v,u}_{\infty}$} (u);
\draw[edge, blue] (P) --
    node[capacity label, below, pos=0.36]
    {$\lrd{P,u}_{\infty}$} (u);

\draw[edge, red] (u) --
    node[capacity label, above, pos=0.55]
    {$\lrd{u,t}_{r_u}$} (t);
\end{tikzpicture}
\end{figure}

In $N$, the maximum flow from $s$ to $t$ equals $\min\{\ell(\S),m\}=\ell(\S)$ by the previous construction. 
Each infinite capacity can be replaced by $\ell(\S)$
without changing the feasible flows.
Then, since all edge capacities are integers, there exists an optimal solution $\{x^*_e\}_{e\in E_N}$ such that $x^*_e\in\mathbb{N}$ for each $e$.
This solution implies that, for each trivial path $v\in J_\S$ (resp., each path $P\in \P'_\S$) and depot $u\in D$, we match $x^*_{\lrd{v,u}}$ (resp., $x^*_{\lrd{P,u}}$) vehicles from $u$ to $v$ (resp., $P$).
Hence, this yields an assignment from the vehicles in $L$ to the paths in $J_\S\cup \P'_\S$: for each trivial path $v\in J_\S$ and depot $u\in D$, we obtain $x^*_{\lrd{v,u}}$ atomic tours $uvu$, where each tour $T$ satisfies $\lambda_{v,T}=Q$; 
for each path $P\in \P'_\S$ and depot $u\in D$ with $x^*_{\lrd{P,u}}=1$, we obtain a single tour $T$ such that $\lambda_{v,T}=\lambda'_{v,P}$ for each $v\in J(P)$, by doubling all edges in $E(P)\cup\{v'u\}$ and then shortcutting, where $v'=\arg\min_{w\in J(P)}c(w,u)$.

The details of \textsc{Transform} are provided in Algorithm~\ref{alg2}.

\begin{algorithm}[!t]
\small
\caption{\textsc{Transform}: Transforming a cycle cover in $G$ into a solution to the MD-SDVRP}
\label{alg2}
\textbf{Input:} A cycle cover $\S$ in $G$ with $\ell(\S)\leq m$. \\
\textbf{Output:} A feasible solution $\T$.
\begin{algorithmic}[1]

\State Update $\S$ into a cycle cover in $G[J]$ by shortcutting, and initialize $J_\S\coloneq \emptyset$, $\P'_\S\coloneq \emptyset$, and $\T\coloneq\emptyset$.\label{alg2s1}

\For{each cycle $C\in\S$}\label{alg2s2}
\State Let $J_C=\{v\in J(C)\mid q_v\geq Q\}$, and $q'_v\coloneq q_v-\floor{\frac{q_v}{Q}}\cdot Q$ for each $v\in J(C)$. \label{alg2s3}
\State Obtain a set of paths $\P'_C$ with the demand assignment $\lambda'_{v,P}$ from $C$ by using $q'_v$ and the path extraction method in~\citep{journals/ijoc/LaiXXD23}, and then update $J_\S\coloneq J_\S\cup J_C$ and $\P'_\S\coloneq\P'_\S\cup \P'_C$.\label{alg2s4}
\EndFor

\State Construct the network $N=(V_N=\{s,t\}\cup J_\S\cup\P'_\S\cup D,E_N)$ as an MCMFP instance.\label{alg2s6}

\State Compute an optimal solution $\{x^*_e\}_{e\in E_N}$ to the MCMFP instance $N$ using the algorithm in~\citep{DBLP:journals/ior/Orlin93}.\label{alg2s7}

\For{each $u\in D$}\label{alg2s8}
\State For each $v\in J_\S$ with $x^*_{\lrd{v,u}}>0$, record $x^*_{\lrd{v,u}}$ atomic tours $uvu$, each tour $T$ satisfying $\lambda_{v,T}=Q$, and add them to $\T$. 
Moreover, for each $P\in \P'_\S$ with $x^*_{\lrd{P,u}}=1$, construct a single tour $T$ with $\lambda_{v,T}=\lambda'_{v,P}$ for each $v\in J(P)$ by doubling all edges in $E(P)\cup\{v'u\}$ and shortcutting ($v'=\arg\min_{w\in J(P)}c(w,u)$), and add $T$ to $\T$.
\EndFor
\State \Return $\T$.\label{alg2s11}

\end{algorithmic}
\end{algorithm}

\subsection{The Analysis}

Consider an arbitrary spanning even-degree (multi-)graph $H$ on $V$, where every component contains a depot. Let $\HH$ denote its component set.
We first prove the following structural property.

\begin{lemma}\label{lem:weighted-cover}
The threshold $\tau$ in line~\ref{alg:main-threshold} of \textsc{Alg.1} satisfies $\tau\leq\frac{\varepsilon c(\HH)}{k}$.
Moreover, for the component set $\B$, there exists a partition $\mathfrak{B}$ of $\B$ such that $\size{\mathfrak{B}}\leq k$, $V(\B_i)\cap D\neq\emptyset$ for each $\B_i\in \mathfrak{B}$, and the corresponding cycle cover $\S$ is a good component set w.r.t.\ $\HH$ and $c(\S)\leq\alpha c(\HH)+2\alpha(k-1)\tau$.
\end{lemma}
\begin{proof}
First, we bound the value $\tau$. Since every component of $\HH$ contains a depot, $|\HH|\leq k$. 
Let $\tau'$ be the largest value in $\lrc{0}\cup\lrc{c(e)}_{e\in E}$ such that $\tau'\leq \frac{\varepsilon c(\HH)}{k}$.
Then, every edge of $E(\HH)$ with cost greater than $\tau'$ has cost greater than $\frac{\varepsilon c(\HH)}{k}$. 
There are fewer than $\frac{k}{\varepsilon}$ such edges, counting their multiplicities, as otherwise their total cost would exceed $c(\HH)$.
Therefore, after deleting all these edges, the remaining graph has at most $\size{\HH}+\ceil{\frac{k}{\varepsilon}}\leq K$ components. All its edges belong to $G_{\tau'}$, so $G_{\tau'}$ also has at most $K$ components.
Since $\tau$ is the smallest value in the set such that $G_\tau$ has at most $K$ components, we obtain $\tau\leq\tau'\leq\frac{\varepsilon c(\HH)}{k}$.

Next, we construct a partition $\mathfrak{B}$ of $\B$ satisfying the three properties in the lemma.

Consider the corresponding component set $\S_{E_\tau\cup E(\HH)}$.
Each component in $\S_{E_\tau\cup E(\HH)}$ corresponds to a set of components in $\B$, so it defines a partition $\mathfrak{B}$ of $\B$.
Clearly, $|\mathfrak{B}|=|\S_{E_\tau\cup E(\HH)}|\leq |\HH|\leq k$.
Since every component of $\HH$ contains a depot, every part of $\mathfrak{B}$ contains a depot.
Thus, $V(\B_i)\cap D\neq\emptyset$ for each $\B_i\in \mathfrak{B}$.
Moreover, every component of $\HH$ is contained in a single
component of $\S_{E_\tau\cup E(\HH)}$, and hence all its vertices
belong to one cycle in $\S$. Thus, $\S$ is good with respect to $\HH$.

It remains to bound the cost of the cycle cover $\S=\lrc{C_i}_{\B_i\in \mathfrak{B}}$ with respect to $\mathfrak{B}$.

By definition, for each part $\B_i\in \mathfrak{B}$, there exists a corresponding component $S_i\in\S_{E_\tau\cup E(\HH)}$ and a corresponding component set $\HH_i\subseteq\HH$ such that $V(\B_i)=V(S_i)=V(\HH_i)$.
Therefore, the components in $\HH_i$ can be connected using $|\HH_i|-1$ edges of $E_\tau$.  
Since each edge of $E_\tau$ has cost at most $\tau$, by using two copies of these edges and using the edges in $E(\HH_i)$, an Eulerian graph on $V(\B_i)$ can be obtained with cost at most $c(\HH_i)+2(|\HH_i|-1)\tau$.
Since $C_i$ is an $\alpha$-approximate TSP solution in $G[V(\B_i)]$, we obtain $c(C_i)\leq \alpha\cdot \lra{c(\HH_i)+2(|\HH_i|-1)\tau}$.
Therefore,
\[
c(\S) = \sum_{\B_i\in\mathfrak{B}}c(C_i)\leq \alpha \sum_{\B_i\in\mathfrak{B}}\lra{c(\HH_i)+2(|\HH_i|-1)\tau} = \alpha\lrA{c(\HH) + 2(|\HH|-|\mathfrak{B}|)\tau}\leq \alpha c(\HH) + 2\alpha(k-1)\tau,
\]
where the last inequality follows from $\size{\HH}\leq k$.
\end{proof}

\begin{lemma}\label{flowtime}
Given a cycle cover $\S$ with $\ell(\S)\leq m$, \textsc{Transform} computes a feasible solution to the MD-SDVRP in $\O(\size{V}^4\log \size{V})$ time, and the solution uses $\ell(\S)$ vehicles.
\end{lemma}
\begin{proof}
We first establish the feasibility of the solution produced by \textsc{Transform}. 
In the constructed MCMFP instance $N$, the maximum flow from $s$ to $t$ exactly equals $\min\{\ell(\S),m\} = \ell(\S)$. Since all edge capacities are integral, an optimal integral flow of value $\ell(\S)$ can be found in polynomial time~\citep{DBLP:journals/ior/Orlin93}. This flow guarantees that all paths in $J_\S\cup \P'_\S$ are successfully satisfied using exactly $\ell(\S)$ vehicles, yielding a feasible solution to the MD-SDVRP.

Next, we analyze the running time of \textsc{Transform}. 
Shortcutting the input cycle cover $\S$ (line~\ref{alg2s1}) takes $\O(\size{V})$ time. 
The computation of sets $J_\S$ and $\P'_\S$ (lines~\ref{alg2s2}--\ref{alg2s4}) is dominated by path extraction. The trivial path set $J_\S$ is extracted in $\O(\size{J})$ time. For each $C\in\S$, since the residual demands satisfy $q'_v<Q$, we have $\size{\P'_C}=\ceil{\frac{\sum_{v\in J(C)}q'_v}{Q}}\leq \size{J(C)}$. Consequently, $\size{\P'_\S}\leq \size{J}$, meaning the path extraction for all cycles also completes in $\O(\size{J})$ time. Thus, lines~\ref{alg2s2}--\ref{alg2s4} take $\O(\size{V})$ time overall.

Constructing the network $N$ (line~\ref{alg2s6}) involves generating vertices and edges based on $J_\S$, $\P'_\S$, and $D$. Since $\size{V_N} = \O(\size{J} + \size{D}) = \O(\size{V})$ and $\size{E_N} = \O(\size{J}\size{D}) = \O(\size{V}^2)$, the network construction takes $\O(\size{V}^2)$ time. 
Solving the MCMFP instance using the algorithm in~\citep{DBLP:journals/ior/Orlin93} (line~\ref{alg2s7}) requires $\O(\size{E_N}(\size{E_N}+\size{V_N}\log\size{V_N})\log \size{V_N})$ time, which is bounded by $\O(\size{V}^4\log \size{V})$. 
Finally, constructing the resulting tours (lines~\ref{alg2s8}--\ref{alg2s11}) takes $\O(\size{V}^2)$ time. 

Hence, \textsc{Transform} runs in $\O(\size{V}^4\log \size{V})$ time.
\end{proof}

Note that when $\size{L}$ and all edge costs are polynomially bounded, the running time of \textsc{Transform} can be reduced to $\size{V}^{2+o(1)}$ by applying the algorithm of~\citep{DBLP:journals/jacm/ChenKLPGS25} to the MCMFP instance.

\begin{lemma}\label{flowcost}
When the input cycle cover $\S$ is a good set of components, the solution $\T$ computed by \textsc{Transform} satisfies $c(\T)\leq 2c(\S)+\OPT$.
\end{lemma}
\begin{proof}
By Property~\ref{good}, we have $\ell(\S)\leq m$. Thus, Lemma~\ref{flowtime} ensures that the solution $\T$ computed by \textsc{Transform} is feasible for the MD-SDVRP.

Let $\fOPT=\sum_{e\in E_N}x^*_e\cdot c(e)$ be the optimal cost of the MCMFP instance in line~\ref{alg2s7} of \textsc{Transform}. Based on the tour construction in lines~\ref{alg2s8}--\ref{alg2s11}, we have
\begin{align*}
c(\T)&\leq 2\lrA{c(J_\S)+c(\P'_\S)+\fOPT} = 2\lrA{c(\P'_\S)+\fOPT}\leq 2\lrA{c(\S)+\fOPT},
\end{align*}
where the equality holds because the trivial paths in $J_\S$ have zero cost, and the second inequality follows from the cost guarantee of the path extraction method~\citep{journals/ijoc/LaiXXD23}. 

It thus suffices to prove the following claim.

\begin{claim}\label{clm1}
It holds that $2\cdot\fOPT\leq\OPT$.
\end{claim}
\begin{claimproof}
We use the optimal MD-SDVRP solution $\T^*$ to construct a feasible flow $\{x_e\}_{e\in E_N}$ for $N$ with a total cost of at most $\frac{1}{2}\cdot\OPT$. The optimality of $\{x^*_e\}_{e\in E_N}$ then implies $2\cdot\fOPT\leq \OPT$.

First, we route flow from the source $s$: we set $x_{\lrd{s,v}}=\floor{\frac{q_v}{Q}}$ for each $v\in J_\S$ and $x_{\lrd{s,P}}=1$ for each $P\in\P'_\S$. Next, we construct valid flows from the paths in $J_\S\cup\P'_\S$ to the depots in $D$ (i.e., mapping paths to the vehicles in $L$).

Since $\S$ is a good component set, there exists a partition $\mathfrak{T}^*=\{\T^*_C\}_{C\in\S}$ of $\T^*$ such that for every $C\in\S$, we have $\T^*_C\subseteq\T^*$ and $J(C)=J(\T^*_C)$. Let $L_C$ denote the set of vehicles utilized in $\T^*_C$. Because $L_C\cap L_{C'}=\emptyset$ for any distinct $C,C'\in\S$, we can independently match the paths in $J_C\cup\P'_C$ to the vehicles in $L_C$ for each component $C$.

For any tour $T\in \T^*_C$, let $\tau_T$ be the associated vehicle and $u_T$ be its starting depot. To match the demand properly, we replace each trivial path $v\in J_C$ (which requires a flow of $\floor{\frac{q_v}{Q}}$) with exactly $\floor{\frac{q_v}{Q}}$ unit-load trivial paths, denoted $v_1,\dots,v_{\floor{\frac{q_v}{Q}}}$, setting $c(v_i,u)=c(v,u)$ for all $u\in D$. Let $J'_C$ be the set of these new unit-load trivial paths, and define $\P''_C=J'_C\cup\P'_C$. Note that $\size{\P''_C}=\ell(C)$.

Since $\T^*_C$ serves all customers in $J(C)$, there is a valid demand assignment $\lambda''_{v,P,T}$ such that each tour $T\in\T^*_C$ serves $\lambda''_{v,P,T}$ demand for customer $v \in J(P)$ for each $P\in\P''_C$. Thus, $\T^*_C$ fully satisfies all paths in $\P''_C$.

We construct a bipartite graph with $\P''_C$ as the left vertex set and $L_C$ as the right vertex set. We add an edge between $P\in\P''_C$ and $\tau_T\in L_C$ if and only if tour $T$ serves a strictly positive demand for path $P$ (i.e., $\lambda''_{v,P,T}>0$ for some $v\in J(P)$). For any subset $\P\subseteq\P''_C$, let $L(\P)\subseteq L_C$ denote its neighborhood in this bipartite graph.

By the property of the path extraction method~\citep{journals/ijoc/LaiXXD23}, at most one path in any subset $\P$ has a demand strictly less than $Q$; all other paths have a demand of exactly $Q$. Thus, the total demand of the paths in $\P$ is strictly greater than $(\size{\P}-1)Q$. Since the vehicles in $L(\P)$ must collectively serve this demand and each vehicle has a capacity of $Q$, their total capacity $\size{L(\P)}\cdot Q$ must be strictly greater than $(\size{\P}-1)Q$. This implies $\size{L(\P)} \geq \size{\P}$. By Hall's Marriage Theorem~\citep{books/springer/Schrijver03}, there exists a matching that assigns each path in $\P''_C$ to a distinct vehicle in $L_C$.

We utilize this matching to route our network flow. For each $v\in J_C$, its corresponding $\floor{\frac{q_v}{Q}}$ unit-load paths in $J'_C$ are matched to $\floor{\frac{q_v}{Q}}$ distinct vehicles in $L_C$. For each matched vehicle $\tau_T\in L_C$, we route one unit of flow from $v$ to its depot $u_T$. 
By the triangle inequality, $c\lrd{v,u_T}\leq \frac{1}{2}c(T)$.
Similarly, each path $P\in\P'_C$ is matched to a single vehicle $\tau_T\in L_C$, and we route one unit of flow from $P$ to $u_T$, incurring a cost of $c\lrd{P,u_T}\leq \frac{1}{2}c(T)$.

Summing over all matched paths, the flow cost for component $C$ is bounded by $\sum_{\tau_T\in L_C}\frac{1}{2}c(T) = \sum_{T\in \T^*_C}\frac{1}{2}c(T)$. Finally, summing over all components $C\in\S$ yields a total flow cost of at most $\sum_{C\in\S}\sum_{T\in \T^*_C}\frac{1}{2}c(T) = \frac{1}{2}\cdot\OPT$. This completes the claim.
\end{claimproof}

This finishes the proof of the lemma.
\end{proof}

By Lemmas~\ref{lem:weighted-cover}, \ref{flowtime}, and \ref{flowcost}, we obtain the following theorem.

\begin{theorem}\label{res:weighted}
For any constant $\varepsilon>0$, the MD-SDVRP admits a $\lra{2\alpha+1+\varepsilon}$-approximation algorithm with running time $2^{\O(\frac{k\log k}{\varepsilon})}\cdot\O(|V|f(|V|)+\size{V}^4\log \size{V})$, where $f(|V|)$ denotes the running time of the given $\alpha$-approximation algorithm for the TSP.
\end{theorem}
\begin{proof}
Let $H=G_{E(\T^*)}$, so $H$ is a spanning even-degree (multi-)graph on $V$, every component of $H$ contains a depot, and $c(H)=\OPT$.
By line~\ref{alg:main-partition} of \textsc{Alg.1}, Property~\ref{good}, and Lemma~\ref{lem:weighted-cover}, there exists a feasible partition of $\B$ whose cycle cover $\S$ yields a good component set. Then, by Lemma~\ref{flowcost},
\[
c(\T_f)\leq2c(\S)+\OPT\leq(2\alpha+1)\OPT+4\alpha(k-1)\tau\leq\left(2\alpha+1+4\alpha\varepsilon\frac{k-1}{k}\right)\OPT= (2\alpha+1+\O(\varepsilon))\OPT.
\]

Next, we analyze the running time. 
Recall that there are $k^K\leq2^{\O(\frac{k\log k}{\varepsilon})}$ possible partitions.
For each partition, there are at most $k\leq |V|$ TSP calls, and \textsc{Transform} takes $\O(\size{V}^4\log \size{V})$ time by Lemma~\ref{flowtime}.
Therefore, the total running time is $2^{\O(\frac{k\log k}{\varepsilon})}\cdot\O(|V|f(|V|)+\size{V}^4\log \size{V})$.

By rescaling $\varepsilon$ to $\frac{\varepsilon}{4\alpha}$, we obtain an approximation ratio of $2\alpha+1+\varepsilon$. 
We may assume that $\alpha$ is a constant, so this rescaling does not change the asymptotic running time.
\end{proof}

\subsection{An Application to the MD-TSP}\label{sec:mdtsp}
The MD-TSP is a special case of the MD-SDVRP with $Q=\infty$. By shortcutting, we may assume that each depot uses at most one vehicle and each customer belongs to exactly one tour.
We may apply the component construction in \textsc{Alg.1}. For each partition $\mathfrak{B}$, the corresponding cycle cover $\S$ can be directly transformed into a feasible solution by shortcutting.
By Lemma~\ref{lem:weighted-cover}, one such solution has cost at most $\alpha\OPT+2\alpha(k-1)\tau\leq(\alpha+\O(\varepsilon))\OPT$.
Thus, by rescaling $\varepsilon$, the framework yields an $(\alpha+\varepsilon)$-approximation algorithm for the MD-TSP.
Since the cost of each cycle depends only on its corresponding part, the running time can be improved via dynamic programming.

\begin{theorem}\label{res:mdtsp}
For any constant $\varepsilon>0$, the MD-TSP admits an $\lra{\alpha+\varepsilon}$-approximation algorithm with running time $2^{\O(\frac{k}{\varepsilon})}\cdot\O(f(|V|)+|V|^2\log|V|)$, where $f(|V|)$ denotes the running time of the given $\alpha$-approximation algorithm for the TSP.
\end{theorem}
\begin{proof}
We first introduce the dynamic programming algorithm.
For each nonempty $\B'\subseteq\B$ containing a depot, compute an $\alpha$-approximate TSP cycle $C_{\B'}$ in $G[V(\B')]$ and let $a(\B')=c(C_{\B'})$; otherwise, let $a(\B')=\infty$.
For each $\B'\subseteq\B$, let $\mathrm{dp}(\B')$ be the minimum total cost of the computed cycles over all partitions of $\B'$.
Set $\mathrm{dp}(\emptyset)=0$. For nonempty $\B'$, we have 
\begin{equation}\label{eq:mdtsp-dp}
\mathrm{dp}(\B')=\min_{\emptyset\neq\B''\subseteq\B'}\lrC{a(\B'')+\mathrm{dp}(\B'\setminus\B'')}.
\end{equation}
Thus, a desired partition $\mathfrak{B}$ of $\B$ can be obtained by computing all values $\mathrm{dp}(\B')$ in nondecreasing order of $\size{\B'}$. 
By construction, each part of $\mathfrak{B}$ must contain a depot, so $\size{\mathfrak{B}}\leq k$.
Thus, by Lemma~\ref{lem:weighted-cover}, the corresponding cycle cover has cost at most $\alpha\OPT+2\alpha(k-1)\tau\leq(\alpha+\O(\varepsilon))\OPT$.

For the running time, there are at most $2^K$ TSP calls, and the total number of transitions in~\eqref{eq:mdtsp-dp} is at most $\sum_{\B'\subseteq\B}2^{\size{\B'}}=3^{\size{\B}}\leq 3^K$. The component construction takes $\O(|V|^2\log|V|)$ time by sorting the edge costs and merging components as edges are added in nondecreasing order of cost.
Thus, the total running time is $2^{\O(\frac{k}{\varepsilon})} \cdot \O(f(|V|)+|V|^2\log|V|)$.
Finally, by rescaling $\varepsilon$ to $\frac{\varepsilon}{2\alpha}$, we obtain an approximation ratio of $\alpha+\varepsilon$ with the same asymptotic running time.
\end{proof}

\section{A Parameterized Approximation Algorithm Related to Vehicle Capacity}\label{sec5}
In this section, we present a parameterized $5$-approximation algorithm (\textsc{Alg.2}) for the MD-SDVRP, whose running time is related to the vehicle capacity and the number of depots. In particular, the algorithm runs in FPT time with a single-exponential dependence on $k$ when $Q=\O(1)$, and in polynomial time when $mQ-\sum_{v\in J}q_v=\O(1)$. 

\subsection{The Algorithm}\label{sec:warmup}
\textsc{Alg.1} relies on enumerating a good component set with a small cost.
\textsc{Alg.2} instead uses a different approach by enumerating the value of $q_u$ for each $u\in D$, where $q_u=(Q-\Lambda^*_u)\bmod Q$, and $\Lambda^*_u$ is the total demand served by vehicles located at $u$ to serve customers in the optimal solution $\T^*$.
The dummy demand $q_u$ makes $\Lambda^*_u+q_u$ divisible by $Q$ without requiring additional vehicles. This allows us to construct a cycle cover whose component demands are divisible by $Q$, so that every extracted path carries $Q$ demand.

Since $\sum_{u\in D}\Lambda^*_u = \sum_{v\in J}q_v$, we have $\sum_{u\in D}q_u \equiv -\sum_{v\in J}q_v \pmod Q$. Thus, it suffices to enumerate $\{q_u\}_{u\in D}$ for only $k-1$ depots, which requires at most $Q^{k-1}$ iterations.
\textsc{Alg.2} constructs a feasible solution in each iteration and returns the best one found across all iterations.
Thus, we assume that $\{q_u\}_{u\in D}$ is given in the subsequent discussion.
In addition, we establish the following property.

\begin{property}\label{depot}
It holds that $\sum_{u\in D}q_u+\sum_{v\in J}q_v\leq mQ$.
\end{property}
\begin{proof}
Since $\T^*$ uses at most $m$ vehicles to serve all customers in $J$, $\sum_{u\in D}\ceil{\frac{\Lambda^*_u}{Q}}\leq m$.
Then, since $\ceil{\frac{\Lambda^*_u}{Q}}=\frac{q_u+\Lambda^*_u}{Q}$ for each $u\in D$ and $\sum_{u\in D}\Lambda^*_u=\sum_{v\in J}q_v$ by definition, $\sum_{u\in D}q_u+\sum_{v\in J}q_v\leq mQ$.
\end{proof}

Given $\{q_u\}_{u\in D}$, \textsc{Alg.2} first treats each depot $u\in D$ as a dummy customer with demand $q_u$, and then utilizes the vehicles in $L$ to serve all customers in $J'\coloneq J\cup D$.
By Property~\ref{depot}, such a solution exists.
Specifically, \textsc{Alg.2} first computes a cycle cover $\S$ in $G$ such that, for each cycle $C\in\S$, the total demand of all customers in $J'(C)$ is divisible by $Q$.
This cycle cover can be computed in $\O(\size{V}^2\log\size{V})$ time by applying the $2$-approximation algorithm of \citet[Section~3, pp.~307--308]{DBLP:journals/siamcomp/GoemansW95} with the proper function $f(S)=1$ if $\sum_{v\in S}q_v\not\equiv0\pmod Q$, and $0$ otherwise.
Next, \textsc{Alg.2} leverages $\S$ to construct a set of tours $\T$ serving all customers in $J'$ by invoking the sub-algorithm \textsc{Transform}.
Note that $\T$ requires $\ell(\S)=\sum_{C\in\S}\ell(C)$ vehicles, where we redefine $\ell(C)=\frac{\sum_{v\in J'(C)}q_v}{Q}$ for all $C\in\S$ (this is guaranteed to be an integer due to the construction of $\S$).
Finally, \textsc{Alg.2} refines $\T$ by shortcutting all dummy customers.

The details of \textsc{Alg.2} are provided in Algorithm~\ref{alg3}.

\begin{algorithm}[!t]
\small
\caption{\textsc{Alg.2}: A parameterized $5$-approximation algorithm for the MD-SDVRP}
\label{alg3}
\textbf{Input:} An MD-SDVRP instance $I=(G,c,q,r,Q)$. \\
\textbf{Output:} A feasible solution $\T_f$.
\begin{algorithmic}[1]
\State Initialize $\mathfrak{T}\coloneq \emptyset$.\label{alg3s1}

\For{each $\{q_u\}_{u\in D}$ with $q_u\in\{0,\dots,Q-1\}$, $\Lambda\leq mQ$, and $\Lambda\bmod Q=0$, where $\Lambda=\sum_{u\in D}q_u+\sum_{v\in J}q_v$}\label{alg3s2}
\State Treat $u$ as a dummy customer with demand $q_u$ for all $u\in D$, and let $J'\coloneq J\cup D$.

\State Apply the algorithm in~\citep{DBLP:journals/siamcomp/GoemansW95} to compute a cycle cover $\S$ in $G$ such that for each $C\in\S$, the total demand of all customers in $J'(C)$ is divisible by $Q$.\label{alg3s4}

\State Call \textsc{Transform} on $\S$ to obtain a set of tours $\T$, which uses $\ell(\S)$ vehicles and satisfies all customers in $J'$, then update $\T$ by shortcutting all dummy customers, and let $\mathfrak{T}\coloneq\mathfrak{T}\cup\{\T\}$.\label{alg3s5}

\EndFor\label{alg3s6}
\State \Return $\T_f=\arg\min_{\T\in\mathfrak{T}}c(\T)$.
\end{algorithmic}
\end{algorithm}

\subsection{The Analysis}
\begin{lemma}\label{alg3lm1}
When $\{q_u\}_{u\in D}$ is correctly enumerated in line~\ref{alg3s2} of \textsc{Alg.2}, the cycle cover $\S$ in line~\ref{alg3s4} satisfies $c(\S)\leq 2\cdot\OPT$.
\end{lemma}
\begin{proof}
First, when $\{q_u\}_{u\in D}$ is correctly enumerated, the proof of Property~\ref{depot} implies that the optimal solution $\T^*$ serving all customers in $J$ can also be used to serve all customers in $J'$, simply by adjusting its demand assignment.

Then, for each $u\in D$, all vehicles departing from $u$ serve a total demand of $\Lambda^*_u+q_u$ in $\T^*$, which is divisible by $Q$ by definition.
Hence, in the component set $\S^*$ formed by $E(\T^*)$, each component $C\in\S^*$ satisfies the condition that the total demand of all customers in $J'(C)$ is divisible by $Q$.
Thus, by shortcutting $\S^*$, one can obtain a cycle cover $\S'$ in $G$ such that for each $C\in\S'$, the total demand of all customers in $J'(C)$ is divisible by $Q$. By the triangle inequality, $c(\S')\leq c(\S^*)=\OPT$.

Since \textsc{Alg.2} applies the $2$-approximation algorithm in~\citep{DBLP:journals/siamcomp/GoemansW95} to compute $\S$ in line~\ref{alg3s4}, it follows that $c(\S)\leq 2\cdot c(\S')\leq 2\cdot\OPT$.
\end{proof}

Since $\S$ may not form a good component set, we cannot use Lemma~\ref{flowcost} to analyze the cost of $\T$ in line~\ref{alg3s5} of \textsc{Alg.2}.
Nevertheless, a similar result can be proven using an alternative analytical method.

\begin{lemma}\label{alg3lm2}
When $\{q_u\}_{u\in D}$ is correctly enumerated in line~\ref{alg3s2} of \textsc{Alg.2}, the set of tours $\T$ in line~\ref{alg3s5} utilizes at most $m$ vehicles to serve all customers in $J'$, and $c(\T)\leq 2c(\S)+\OPT$.
\end{lemma}
\begin{proof}
We first demonstrate that when $\{q_u\}_{u\in D}$ is correctly enumerated, the cycle cover $\S$ can be fed into the sub-algorithm \textsc{Transform} to produce such a set of tours $\T$ for the new instance, where each $u\in D$ is additionally assigned a demand of $q_u$.

Since $\S$ is a cycle cover in $G$ ensuring the total demand of all customers in $J'(C)$ is divisible by $Q$ for each $C\in\S$, we have $\ell(\S)=\sum_{C\in\S}\frac{\sum_{v\in J'(C)}q_v}{Q}=\frac{\sum_{u\in D}q_u+\sum_{v\in J}q_v}{Q}\leq m$ by the definition of the values $\{q_u\}_{u\in D}$ and Property~\ref{depot}.
Hence, by Lemma~\ref{flowtime}, \textsc{Transform} can be applied to $\S$ to generate a set of tours $\T$, which utilizes $\ell(\S)\leq m$ vehicles to serve all customers in $J'$.

Next, we analyze the cost of $\T$.

Recall that when \textsc{Transform} is applied to $\S$, it first constructs a set of paths $J'_\S\cup \P_\S$ and then computes a minimum-cost maximum flow in the MCMFP instance $N$ to match all paths in $J'_\S\cup \P_\S$ to the $\ell(\S)$ vehicles in $L$.
Let $\fOPT$ be the optimal cost in $N$. 
By the proof of Lemma~\ref{flowcost}, $c(\T)\leq 2c(\S)+2\fOPT$.

To complete the proof, it suffices to establish the following claim.

\begin{claim}\label{clm3}
It holds that $2\cdot\fOPT\leq\OPT$.
\end{claim}
\begin{claimproof}
Because $\S$ may not be a good component set, we cannot rely on the analysis method from the proof of Claim~\ref{clm1}.
Instead, we will use $\T^*$ to construct a \emph{fractional} maximum flow $\{x_e\}_{e\in E_N}$ for the MCMFP instance $N$ such that $\sum_{e\in E_N}x_e\cdot c(e)\leq \frac{1}{2}\cdot\OPT$.
Recall that since all edge capacities in $N$ are integral, the minimum-cost fractional maximum flow is equal to the minimum-cost integral maximum flow. 
Hence, by the optimality of $\{x^*_e\}_{e\in E_N}$, we deduce that $2\cdot\fOPT\leq \OPT$.

Based on the proof of Lemma~\ref{alg3lm1}, when $\{q_u\}_{u\in D}$ is correctly enumerated, the set of tours $\T^*$ serving all customers in $J$ can also be utilized to serve all customers in $J'$, simply by adjusting its demand assignment.
Thus, there must exist a demand assignment $\lambda'_{v,P,T}$ such that, for each path $P\in J'_\S\cup \P_\S$, each tour $T\in\T^*$ assigns $\lambda'_{v,P,T}$ demand to each customer $v\in J'(P)$. Moreover, $\T^*$ satisfies all paths in $J'_\S\cup \P_\S$.
For any tour $T\in \T^*$, let $\tau_T$ denote the vehicle used, and $u_T$ denote the depot where $\tau_T$ is located.
The fractional flow is constructed as follows:

For each tour $T\in\T^*$, each path $P\in J'_\S\cup\P_\S$, and each customer $v\in J'(P)$, whenever $\lambda'_{v,P,T}>0$, we sent $\frac{\lambda'_{v,P,T}}{Q}$ flow along the network vertices $s, P, u_T, t$.

The total cost of the constructed flow is bounded by:
\begin{align*}
&\sum_{T\in\T^*}\sum_{P\in J'_\S\cup\P_\S}\sum_{v\in J'(P)} \frac{\lambda'_{v,P,T}}{Q}\cdot c\lrd{P,u_T}\leq\sum_{T\in\T^*}\sum_{P\in J'_\S\cup\P_\S}\sum_{v\in J'(P)} \frac{\lambda'_{v,P,T}}{Q}\cdot\frac{1}{2}c(T)\leq\sum_{T\in\T^*}\frac{1}{2} c(T)=\frac{1}{2}\cdot \OPT,
\end{align*}
where the first inequality leverages the fact that when $v\in J'(P)$, $c\lrd{P,u_T}=\min_{w\in J'(P)}c(w,u_T)\leq c(v,u_T)$ holds by definition. Furthermore, when $\lambda'_{v,P,T}>0$, $c(v,u_T)\leq \frac{1}{2}c(T)$ holds due to the triangle inequality. The second inequality follows from $\sum_{P\in J'_\S\cup\P_\S}\sum_{v\in J'(P)}\lambda'_{v,P,T}\leq Q$, as each tour $T\in\T^*$ can fulfill at most $Q$ demand.

Finally, we verify that this flow construction yields a feasible fractional maximum flow in $N$.

First, since each dummy customer $u\in D$ has a demand of at most $Q-1$, the set of trivial paths $J'_\S$ is exactly $\{v\in J\mid q_v\geq Q\}$ by the definition of \textsc{Transform}. Moreover, each trivial path $v\in J'_\S$ has a load of $\floor{\frac{q_v}{Q}}$ and a demand of $\floor{\frac{q_v}{Q}}\cdot Q$.
Similarly, because the total demand of all customers in $J'(C)$ is divisible by $Q$ for each $C\in\S$, each $P\in \P_\S$ has a load of $1$ and a demand of exactly $Q$.

Consider each $P\in J'_\S\cup\P_\S$.
If $P=v$ is a trivial path, it has a demand of $\floor{\frac{q_v}{Q}}\cdot Q$; under the flow construction, the total flow sent from $s$ to $P$ is precisely $\floor{\frac{q_v}{Q}}$.
Otherwise, $P$ has a demand of exactly $Q$, and the total flow sent from $s$ to $P$ equals $1$.

Consequently, a total flow of $\sum_{v\in J'_\S} \floor{\frac{q_v}{Q}} + \sum_{P\in \P_\S} 1 = \ell(\S)$ is routed from $s$ to all paths in $J'_\S\cup \P_\S$, and ultimately to $t$. For every path $P\in J'_\S\cup\P_\S$, the capacity of $\lrd{s,P}$ exactly matches the assigned flow, ensuring none of these capacities are violated.
Therefore, the constructed flow is a valid maximum $s$-$t$ flow.

Next, since every edge of the form $\lrd{P,u}$ for $P\in J'_\S\cup\P_\S$ and $u\in D$ has infinite capacity, their capacity constraints are trivially respected. 
It remains only to verify the capacity constraints for edges of the form $\lrd{u,t}$ for each $u\in D$.

Consider any depot $u\in D$.
Let $\T^*_u=\{T\in\T^*\mid u_T=u\}$ denote the set of tours in $\T^*$ served by vehicles from $u$.
According to the flow construction, vertex $u$ in the network receives a total incoming flow of 
\[
\sum_{T\in\T^*_u}\sum_{P\in J'_\S\cup\P_\S}\sum_{v\in J'(P)} \frac{\lambda'_{v,P,T}}{Q}\leq \sum_{T\in\T^*_u}1=\size{\T^*_u}\leq r_u,
\]
where the first inequality reflects that each tour $T\in\T^*$ serves at most $Q$ demand, and the final inequality holds because $\T^*$ is a feasible solution serving all customers in $J'$.
Thus, the capacity constraint of $\lrd{u,t}$ is satisfied.
\end{claimproof}
This finishes the proof of the lemma.
\end{proof}

By Lemmas~\ref{flowtime}, \ref{alg3lm1}, and \ref{alg3lm2}, we obtain the following theorem.

\begin{theorem}\label{res3}
For the MD-SDVRP, \textsc{Alg.2} runs in $\O(\min\{Q^{k-1},\binom{mQ-\sum_{v\in J}q_v+k-1}{k-1}\}\cdot\size{V}^4\log \size{V})$ time and achieves an approximation ratio of $5$.
\end{theorem}
\begin{proof}
We first verify that \textsc{Alg.2} correctly enumerates $\{q_u\}_{u\in D}$ in lines~\ref{alg3s2}--\ref{alg3s6}.
It suffices to show that the conditions $\Lambda\leq mQ$ and $\Lambda\bmod Q=0$ in line~\ref{alg3s2} do not constrain the valid enumerations.

Let $\Lambda=\sum_{v\in J'}q_v$.
When $\{q_u\}_{u\in D}$ is correctly enumerated, the definitions guarantee that $\Lambda\leq mQ$ and $\Lambda\bmod Q=0$.
Thus, these conditions do not compromise the completeness of the enumeration.

By Lemmas~\ref{alg3lm1} and \ref{alg3lm2}, \textsc{Alg.2} computes a set of tours $\T$ with $c(\T)\leq 5\cdot\OPT$, while using at most $m$ vehicles to serve all customers in $J'$.
Subsequently, by line~\ref{alg3s5} and the triangle inequality, shortcutting all dummy customers yields a valid $5$-approximate solution for the MD-SDVRP.

We now analyze the running time.

As previously established, since $\sum_{u\in D}q_u \equiv -\sum_{v\in J}q_v \pmod Q$, it suffices to enumerate $\{q_u\}_{u\in D}$ for only $k-1$ depots, which yields at most $Q^{k-1}$ iterations.
On the other hand, when the residual capacity gap $mQ-\sum_{v\in J}q_v$ is small, the search space for feasible choices of $\{q_u\}_{u\in D}$ is significantly reduced. 
Specifically, because $\sum_{u\in D}q_u+\sum_{v\in J}q_v\leq mQ$ as shown in Property~\ref{depot}, the number of feasible assignments for $\{q_u\}_{u\in D}$ is additionally bounded by the number of ways to distribute $mQ-\sum_{v\in J} q_v$ identical balls into $k$ distinct boxes (permitting empty boxes), which evaluates to $\binom{mQ-\sum_{v\in J}q_v+k-1}{k-1}$. 
Thus, the total number of iterations is at most $\min\{Q^{k-1},\binom{mQ-\sum_{v\in J}q_v+k-1}{k-1}\}$.

Within each iteration, the algorithm of~\citep{DBLP:journals/siamcomp/GoemansW95} in line~\ref{alg3s4} operates in $\O(\size{V}^2\log\size{V})$ time, and the sub-algorithm \textsc{Transform} in line~\ref{alg3s5} executes in $\O(\size{V}^4\log \size{V})$ time by Lemma~\ref{flowtime}.
Thus, the overall running time is $\O(\min\{Q^{k-1},\binom{mQ-\sum_{v\in J}q_v+k-1}{k-1}\}\cdot\size{V}^4\log \size{V})$.
\end{proof}

\section{A Bi-Factor Approximation Framework}\label{sec6}
In this section, we establish a bi-factor approximation framework (\textsc{Alg.3}) for the MD-SDVRP.
By treating any bi-factor approximation algorithm for the CMD-VRP as a black box, the framework provides a generic bi-factor reduction.
By instantiating this framework and refining the structural analysis, we achieve a bi-factor approximation ratio of $(6+2/\varepsilon+2\varepsilon/(1+\varepsilon), 1+\varepsilon)$ in polynomial time.
Furthermore, we show that the core routing extraction technique within \textsc{Alg.3} can be adapted to yield an $(\alpha+1-\varepsilon_\alpha)$-approximation algorithm for the SDVRP.

\subsection{The Algorithm}
\textsc{Alg.3} is motivated by a structural connection between the MD-SDVRP and the CMD-VRP.

Recall that in the CMD-VRP, each depot is equipped with an unlimited number of vehicles, but each depot $u\in D$ is assigned a \emph{capacity} of $f(u)\in\mathbb{R}_{\geq0}$.
The goal is to compute a minimum-cost set of tours to serve all customers such that, for each $u$, the total customer demand served by vehicles originating from $u$ is at most $f(u)$.
Let $\OPT'$ denote the optimal cost of the CMD-VRP.

Given an MD-SDVRP instance $I$, \textsc{Alg.3} first constructs a corresponding CMD-VRP instance $I'$ by setting $f(u)=r_u\cdot Q$ for each $u\in D$.
Since any feasible solution for $I$ is feasible for $I'$, $\OPT'\leq \OPT$.
Suppose we are given a bi-factor $(\rho, \sigma)$-approximation algorithm for the CMD-VRP, where $\rho \geq 1$ bounds the cost and $\sigma \geq 1$ parameterizes the capacity violation.
By applying this black box to $I'$, \textsc{Alg.3} computes a set of tours $\T'$ with $c(\T')\leq \rho\cdot \OPT'$, while the total demand served from each depot $u$ is at most $f(u) + (\sigma-1) \cdot Q = (r_u+\sigma-1)\cdot Q$.

Because $\T'$ may deploy more than $r_u$ vehicles from a given depot $u\in D$, \textsc{Alg.3} merges and splits $\T'$ to restore the vehicle count constraint.
Consider any depot $u\in D$, and let $\T'_u$ denote the set of tours in $\T'$ that originate from $u$. 
Suppose these tours serve $\lambda'_v$ demand for each $v\in J$.
The union of tours in $\T'_u$ forms a connected component, which is Eulerian. 
By shortcutting this Eulerian graph, we extract a cycle $C_u$ traversing $u$ and all customers $v$ with $\lambda'_v>0$.
Then, by applying the modified cycle-splitting procedure (Lemma~\ref{sdvrp}) to $C_u$, \textsc{Alg.3} constructs a new set of tours $\T_u$ satisfying $\lambda'_v$ demand for each $v\in J$, where each newly constructed tour serves at most $\sigma \cdot Q$ demand.

As we will formally prove, this procedure guarantees $\size{\T_u}\leq r_u$.
Consequently, $\bigcup_{u\in D} \T_u$ forms a valid bi-factor solution for the MD-SDVRP, violating the individual vehicle capacity by a factor of at most $\sigma$. The description of \textsc{Alg.3} is provided in Algorithm~\ref{alg5}.

\begin{algorithm}[!t]
\small
\caption{\textsc{Alg.3}: A bi-factor approximation framework for the MD-SDVRP}
\label{alg5}
\textbf{Input:} An MD-SDVRP instance $I=(G,c,q,r,Q)$. \\
\textbf{Output:} A set of tours $\T$.
\begin{algorithmic}[1]
\State Initialize $\T\coloneq \emptyset$. Construct a CMD-VRP instance $I'$ by setting $f(u)\coloneq r_u\cdot Q$ for each $u\in D$, and obtain a set of tours $\T'$ by applying a bi-factor $(\rho, \sigma)$-approximation algorithm for the CMD-VRP.\label{alg5s1}

\For{each $u\in D$ with $\T'_u \neq \emptyset$, where $\T'_u$ is the set of tours in $\T'$ using vehicles from $u$}\label{alg5s2}
\State Let $\lambda'_v$ be the demand served by tours in $\T'_u$ for each $v\in J$. 
Obtain a cycle $C_u$ containing $u$ and all customers $v$ with $\lambda'_v>0$ by shortcutting the Eulerian graph formed by $\T'_u$. Apply the procedure in Lemma~\ref{sdvrp} to $C_u$ to obtain a set of tours $\T_u$ satisfying $\lambda'_v$ demand for each $v\in J$, where each tour serves at most $\sigma\cdot Q$ demand. Update $\T\coloneq\T\cup\T_u$.\label{alg5s3}

\EndFor\label{alg5s4}
\State \Return $\T$.
\end{algorithmic}
\end{algorithm}

\subsection{The Analysis}
\begin{lemma}\label{sdvrp}
Given a cycle $C_u=uv_1\dots v_tu$, where each customer $v_i$ has demand $\lambda'_i$, there exists an $\O(t^2)$-time algorithm that computes a set of tours $\T$ for vehicles of capacity $Q'$ from depot $u$ to serve all customers, where $\size{\T}=\ceil{\frac{\sum_{i=1}^{t}\lambda'_i}{Q'}}$ and $c(\T)\leq c(C_u)+\frac{\sum_{i=1}^{t}\lambda'_i\cdot 2c(u,v_i)}{Q'}$.
\end{lemma}
\begin{proof}
The classical ITP algorithm~\citep{journals/moor/HaimovichK85} guarantees the desired cost bound. 
However, a naive implementation of ITP requires $\O(Q't)$ time, which is pseudo-polynomial. 
While a dynamic programming~(DP) approach~\citep{journals/mp/BlauthTV23} can optimize the partition in $\O(t^2)$ time, it may output more than $\ceil{\frac{\sum_{i=1}^{t}\lambda'_i}{Q'}}$ tours, rendering it infeasible for the SDVRP vehicle constraints.

To overcome this without resorting to a highly complex DP modification, we adapt the previous path extraction method of~\citet{journals/ijoc/LaiXXD23} to achieve both the polynomial $\O(t^2)$ running time and the tour count bound. 

We add a dummy customer $v_0$ at $u$ with demand $\lambda'_0\coloneq (Q'-\sum_{i=1}^{t}\lambda'_i)\bmod Q'$.
Then, the total demand $\sum_{i=0}^{t}\lambda'_i$ is divisible by $Q'$, and $\frac{\sum_{i=0}^{t}\lambda'_i}{Q'}=\ceil{\frac{\sum_{i=1}^{t}\lambda'_i}{Q'}}$.
It suffices to construct $\frac{\sum_{i=0}^{t}\lambda'_i}{Q'}$ tours to serve all customers in $\{v_0,\dots,v_t\}$.

We first preprocess each $v_i$ with $\lambda'_i\geq Q'$ by extracting $\floor{\frac{\lambda'_i}{Q'}}$ tours $uv_iu$, and let $\lambda''_i\coloneq \lambda'_i-\floor{\frac{\lambda'_i}{Q'}}\cdot Q'$.
Each such tour serves exactly $Q'$ demand.
Let $\T'$ denote the set of these atomic tours.
It remains to serve the residual demands $\lambda''_i$.

By shortcutting all $v_i$ with $\lambda''_i=0$ from $C_u$, we obtain a cycle $C'_u=v_{p_0}v_{p_1}\dots v_{p_{t'}}v_{p_0}$, where each $v_{p_i}$ has demand $\lambda''_{p_i}$.
Let $u_i\coloneq v_{p_i}$ and $\lambda_i\coloneq \lambda''_{p_i}$.
Then $C'_u=u_0u_1\dots u_{t'}u_0$, $c(C'_u)\leq c(C_u)$ by the triangle inequality, and $\sum_{i=0}^{t'}\lambda_i=\sum_{i=0}^{t}\lambda''_i$.

We choose an initial vertex $u_i$ and apply the path extraction method of~\citep{journals/ijoc/LaiXXD23} to $C'_u$, obtaining a set of paths $\P_i$ with a demand assignment.
Since $\sum_{i=0}^{t'}\lambda_i$ is divisible by $Q'$, each path in $\P_i$ carries $Q'$ demand.
For each $P\in\P_i$, we form a tour $T_P$ by connecting its endpoints to the depot $u$.
Let $\T_i \coloneq \{T_P \mid P\in\P_i\}$ denote the resulting set of tours.

There are $t'+1=\O(t)$ choices of the initial vertex, yielding tour sets $\mathfrak{T}\coloneq\{\T_0,\dots,\T_{t'}\}$.
We select the minimum-cost one $\T'' \coloneq \arg\min_{\T_i\in\mathfrak{T}} c(\T_i)$, and output $\T\coloneq \T'\cup\T''$.
Thus, the computation of $\T$ takes $\O(t^2)$ time, dominated by the computation of $\T''$.
Moreover, $\size{\T}=\size{\T'}+\size{\T''}=\sum_{i=0}^{t}\floor{\frac{\lambda'_i}{Q'}}+\frac{\sum_{i=0}^{t}\lambda''_i}{Q'}=\frac{\sum_{i=0}^{t}\lambda'_i}{Q'}=\ceil{\frac{\sum_{i=1}^{t}\lambda'_i}{Q'}}$.

Hence, it remains to prove that $c(\T)\leq c(C'_u)+\frac{\sum_{i=1}^{t}\lambda'_i\cdot 2c(u,v_i)}{Q'}$.

Since $c(\T')=\sum_{i=0}^{t}\floor{\frac{\lambda'_i}{Q'}}\cdot 2c(u,v_i)$ and $v_0=u$, it suffices to prove  $c(\T'')\leq c(C'_u)+\frac{\sum_{i=0}^{t'}\lambda_{i}\cdot 2c(u,u_{i})}{Q'}$.

Let $C=\lra{u^1_0...u^{\lambda_0}_0}\lra{u^1_1...u^{\lambda_1}_1}...\lra{u^1_{t'}...u^{\lambda_{t'}}_{t'}}u^1_0$, obtained from $C'_u$ by replacing each $u_i$ with $\lambda_i$ unit-demand copies at the same place.
Relabel $C=w_1w_2...w_{t^*}w_1$, where $t^*=\sum_{i=0}^{t'}\lambda_{i}$.
Then, $\frac{\sum_{i=0}^{t'}\lambda_{i}\cdot 2c(u,u_{i})}{Q'}=\frac{\sum_{i=1}^{t^*}2c(u,w_{i})}{Q'}$, and $c(C)=c(C'_u)$.

Applying the tour partition method to $C$ with all possible initial vertices yields exactly $t^*$ tour sets $\mathfrak{T}'\coloneq\{\tilde{\T}_1,\dots,\tilde{\T}_{t^*}\}$.
Let $\tilde{\T}'\coloneq \arg\min_{\tilde{\T}_i\in\mathfrak{T}'}c(\tilde{\T}_i)$.
By the analysis of ITP~\citep{journals/moor/HaimovichK85}, $c(\tilde{\T}')\leq c(C)+\frac{\sum_{i=1}^{t^*}2c(u,w_{i})}{Q'}$.

Note that $t^*$ may be super-polynomial in $t$, so computing $\tilde{\T}'$ requires pseudo-polynomial time. This explains why we cannot apply this method directly as our algorithm, but we can still utilize its structural properties for the proof.

For each $\T_i\in \mathfrak{T}$, let $\tilde{\T}_{q_i}$ be obtained from $\T_i$ by replacing each $u_i$ with $\lambda_i$ unit-demand copies.
Then $c(\T_i)=c(\tilde{\T}_{q_i})$, and $\tilde{\T}_{q_i}\in \mathfrak{T}'$.
Since $\T''=\arg\min_{\T_i\in\mathfrak{T}} c(\T_i)$, it suffices to show that some $\tilde{\T}_{q_i}$ has cost at most $c(\tilde{\T}')$.

\textbf{Case 1:} There exists a tour of the form $uu^1_i\dots u$ in $\tilde{\T}'$.
Then $\tilde{\T}'$ is precisely the partition obtained with initial vertex $u^1_i$, so $\tilde{\T}_{q_i}=\tilde{\T}'$.

\begin{figure}[t]
\centering
\begin{tikzpicture}[
    scale=0.6,
    dot/.style={circle, fill, inner sep=1.5pt},
    edge/.style={-stealth, thick},
    T1/.style={-stealth, thick, red},
    T2/.style={-stealth, thick, blue},
    T3/.style={-stealth, thick, orange},
    T4/.style={-stealth, thick, green!60!black},
    scale=1.2
]

\node[dot] (u) at (0,0) {};
\node[below right] at (u) {$u$};

\def\R{1.8}

\node[dot] (w1) at (0:\R) {};   \node[right=2pt] at (w1) {$u_{i_1}$};
\node[dot] (w2) at (90:\R) {};  \node[above=2pt] at (w2) {$u_{i_2}$};
\node[dot] (w3) at (180:\R) {}; \node[left=2pt]  at (w3) {$u_{i_3}$};
\node[dot] (w4) at (270:\R) {}; \node[below=2pt] at (w4) {$u_{i_4}$};

\node[dot] (m1) at (45:\R) {};
\node[dot] (m2) at (135:\R) {};
\node[dot] (m3) at (225:\R) {};
\node[dot] (m4) at (315:\R) {};

\draw[T1] (u)  to[bend left=15] (w1);
\draw[T1] (w1) to[bend left=15] (m1);
\draw[T1] (m1) to[bend left=15] (w2);
\draw[T1] (w2) to[bend left=15] (u);
\node[red] at (45:\R+0.45) {$T_1$};

\draw[T2] (u)  to[bend left=15] (w2);
\draw[T2] (w2) to[bend left=15] (m2);
\draw[T2] (m2) to[bend left=15] (w3);
\draw[T2] (w3) to[bend left=15] (u);
\node[blue] at (135:\R+0.45) {$T_2$};

\draw[T3] (u)  to[bend left=15] (w3);
\draw[T3] (w3) to[bend left=15] (m3);
\draw[T3] (m3) to[bend left=15] (w4);
\draw[T3] (w4) to[bend left=15] (u);
\node[orange] at (225:\R+0.45) {$T_3$};

\draw[T4] (u)  to[bend left=15] (w4);
\draw[T4] (w4) to[bend left=15] (m4);
\draw[T4] (m4) to[bend left=15] (w1);
\draw[T4] (w1) to[bend left=15] (u);
\node[green!60!black] at (315:\R+0.45) {$T_4$};

\end{tikzpicture}
\caption{An illustration of a split cycle. Consecutive tours $T_1,\dots,T_4$ form a star-like structure centered at the depot $u$. Each tour contains at least two distinct customers to fully serve exactly $Q'$ demand.}
\label{fig:split_cycle}
\end{figure}

\textbf{Case 2:} No tour of the form $uu^1_i\dots u$ exists in $\tilde{\T}'$.
Then no tour starts at the first unit-demand copy of any customer. Therefore, every tour must start at some $u^j_i$ with $j > 1$, which means every tour shares its first visited customer $u_i$ with the preceding tour.
Since $\lambda_i=\lambda''_i \in (0, Q')$ and every tour in $\tilde{\T}'$ serves exactly $Q'$ demand, no single customer can completely fulfill a tour. Thus, every tour must contain at least two distinct customers. 

If $\size{\tilde{\T}'} = 1$, the total demand is exactly $Q'$, and the single tour serves all customers. 
Such a tour can be re-indexed to start at $u^1_0$ without changing its set of edges or total cost, thereby reducing this scenario to Case 1.

For $\size{\tilde{\T}'}\geq2$, we use the demand-shifting argument of \citet[Properties 1 and 2]{journals/nrl/DrorT90}.
Assume $\tilde{\T}'=\{T_1,\dots,T_h\}$ in cyclic order. Let $u_{i_j}$ be the last customer of $T_{j-1}$ and the first customer of $T_j$, where $T_0=T_h$ and $u_{i_{h+1}}=u_{i_1}$. These shared customers form a \emph{split cycle}, as illustrated in Figure~\ref{fig:split_cycle}.
Let $a_j>0$ be the demand served by $T_j$ at $u_{i_j}$, and set $\delta=\min_j a_j$.
For each $j$, decrease the demand served by $T_j$ at $u_{i_j}$ by $\delta$, and increase that served at $u_{i_{j+1}}$ by $\delta$.
Each tour still serves exactly $Q'$ demand.
At each shared customer $u_{i_j}$, the decrease in the delivery from $T_j$ equals the increase in that from $T_{j-1}$, so its demand remains fully satisfied. All delivery amounts remain nonnegative, and no tour visits
any additional customer. This change moves the starting position of every tour forward by $\delta$ unit-demand copies along the cycle.
For some $j$ with $a_j=\delta$, tour $T_j$ no longer serves $u_{i_j}$ and starts at the first unit-demand copy of the next customer. By shortcutting visits with zero delivery, we obtain another valid partition $\tilde{\T}^*$ with $c(\tilde{\T}^*)\leq c(\tilde{\T}')$.
Thus, $\tilde{\T}^*$ satisfies Case~1, and the claim follows.
\end{proof}

\begin{theorem}\label{res4_gen}
Given a polynomial-time $(\rho, \sigma)$-approximation algorithm for the CMD-VRP with the depot-capacity guarantee stated above, \textsc{Alg.3} operates in polynomial time and achieves a ratio of $(\rho+\rho/\sigma, \sigma)$ for the MD-SDVRP.
\end{theorem}
\begin{proof}
By Lemma~\ref{sdvrp}, the cycle-splitting procedure in \textsc{Alg.3} runs in polynomial time.
Since $\T'$ completely serves all customers by line~\ref{alg5s1}, the resulting solution $\T$ serves all customers' demand. Moreover, each constructed tour serves at most $\sigma\cdot Q$ demand according to lines~\ref{alg5s2}--\ref{alg5s4} and Lemma~\ref{sdvrp}.

By the definition of the black-box algorithm, $c(\T')\leq \rho\cdot \OPT' \leq \rho\cdot \OPT$. To prove the theorem, it suffices to show that $\size{\T_u}\leq r_u$ and $c(\T_u)\leq 2c(\T'_u)$.

Regarding the vehicle count, line~\ref{alg5s3} and Lemma~\ref{sdvrp} ensure that $\size{\T_u}=\ceil{\frac{\sum_{v\in J}\lambda'_v}{\sigma\cdot Q}}\leq \ceil{\frac{(r_u+\sigma-1)\cdot Q}{\sigma\cdot Q}} \leq r_u$, where the first inequality uses the fact that $\T'_u$ serves at most $f(u)+(\sigma-1)\cdot Q=(r_u+\sigma-1)\cdot Q$ total demand by line~\ref{alg5s1}, and the final inequality holds because $\ceil{\frac{r_u+\sigma-1}{\sigma}} = \ceil{\frac{r_u-1}{\sigma}} + 1 \leq r_u-1+1=r_u$ for any integer $r_u \geq 1$ and real $\sigma \geq 1$.

Next, we bound the routing cost. The tours in $\T'_u$ serve $\lambda'_v$ demand for each $v\in J$ by line~\ref{alg5s3}.
Because the cycle $C_u$ is obtained by shortcutting the Eulerian graph formed by $\T'_u$, we can invoke the classical VRP lower bounds on $c(\T'_u)$~\citep{journals/moor/HaimovichK85}:
\begin{equation}\label{vrplb}
c(\T'_u)\geq\max\lrC{\frac{\sum_{v\in J}\lambda'_v\cdot 2c(u,v)}{Q}, c(C_u)}.
\end{equation}
Thus, applying Lemma~\ref{sdvrp} with $Q' = \sigma\cdot Q$, we have $c(\T_u)\leq c(C_u)+\frac{\sum_{v\in J}\lambda'_v\cdot 2c(u,v)}{\sigma\cdot Q} \leq (1+1/\sigma)c(\T'_u)$, where we use the property $\sigma \geq 1$. Summing over all depots yields $c(\T)\leq (1+1/\sigma)c(\T')\leq (\rho+\rho/\sigma)\cdot \OPT$.
\end{proof}

By instantiating this framework with the state-of-the-art bi-factor $(4+2/\varepsilon, 1+\varepsilon)$-approximation algorithm for the CMD-VRP~\citep{journals/eor/HeineDM23}, we immediately obtain a bi-factor approximation ratio of $(4+4/\varepsilon+2/(1+\varepsilon), 1+\varepsilon)$ for the MD-SDVRP in polynomial time.
We can obtain a tighter bi-factor approximation ratio by further exploiting a structural property of the algorithm in~\citep{journals/eor/HeineDM23}: in the computed solution, each tour serves at most $\varepsilon\cdot Q$ demand.

\begin{theorem}\label{cor_refined}
For the MD-SDVRP with any constant $\varepsilon>0$, there exists a polynomial-time bi-factor $(6+2/\varepsilon+2\varepsilon/(1+\varepsilon),1+\varepsilon)$-approximation algorithm.
\end{theorem}
\begin{proof}
Since each tour in $\T'$ serves at most $\varepsilon\cdot Q$ demand~\citep{journals/eor/HeineDM23}, the lower bound in \eqref{vrplb} can be tightened to $c(\T'_u)\geq\max\lrc{\frac{\sum_{v\in J}\lambda'_v\cdot 2c(u,v)}{\varepsilon\cdot Q}, c(C_u)}$.
Thus, by Lemma~\ref{sdvrp} with $Q' = (1+\varepsilon)\cdot Q$, we have $c(\T_u)\leq c(C_u) + \frac{\varepsilon}{1+\varepsilon} \cdot \frac{\sum_{v\in J}\lambda'_v\cdot 2c(u,v)}{\varepsilon\cdot Q} \leq \lra{1+\frac{\varepsilon}{1+\varepsilon}}c(\T'_u) = \frac{1+2\varepsilon}{1+\varepsilon}c(\T'_u)$.

Summing over all depots yields $c(\T)\leq \frac{1+2\varepsilon}{1+\varepsilon}c(\T')$.  
Since $c(\T')\leq (4+2/\varepsilon)\cdot\OPT$~\citep{journals/eor/HeineDM23}, we have $c(\T)\leq \frac{1+2\varepsilon}{1+\varepsilon}\cdot \lra{4+\frac{2}{\varepsilon}}\cdot\OPT=\lra{6+\frac{2}{\varepsilon}+\frac{2\varepsilon}{1+\varepsilon}}\cdot\OPT$.
\end{proof}

\subsection{The Extension to the SDVRP}
The algorithm in \citep{journals/ijoc/LaiXXD23} achieves an approximation ratio of $6-4/k$ for the MD-SDVRP for each fixed $k\geq 2$.
When $k=1$, the MD-SDVRP becomes the SDVRP, and it can be verified that their algorithm achieves an approximation ratio of $4$ with running time $\O(\max\{\size{V}^3,\size{L}^3\})$.

The modified splitting procedure in Lemma~\ref{sdvrp} can also be used to close the gap in approximation ratios between the SDVRP and the splittable VRP~\citep{journals/mp/BlauthTV23}.

\begin{theorem}\label{res4}
The SDVRP admits an approximation ratio of $\alpha+1-\varepsilon_\alpha$.
\end{theorem}
\begin{proof}
Given an SDVRP instance, let $\OPT'$ denote the optimal cost of the corresponding splittable VRP instance. Clearly, $\OPT'\leq\OPT$.
Let $C_\alpha$ denote an $\alpha$-approximate TSP tour in $G$.
The VRP lower bounds~\citep{journals/moor/HaimovichK85} guarantee that $\OPT'\geq \max\{\frac{\sum_{v\in J}q_v\cdot 2c(u,v)}{Q}, \frac{1}{\alpha}c(C_\alpha)\}$.

Fix a sufficiently small constant $\varepsilon>0$.
\citet{journals/mp/BlauthTV23} proved that a TSP tour $C_\varepsilon$ in $G$ can be computed in polynomial time with $c(C_\varepsilon)\leq (1+f(\varepsilon))\cdot \OPT'$ whenever $\frac{\sum_{v\in J}q_v\cdot 2c(u,v)}{Q}\geq (1-\varepsilon)\cdot \OPT'$.
The function $f$ satisfies $f(\varepsilon)\geq 0$ and $\lim_{\varepsilon\rightarrow0^+}f(\varepsilon)=0$.

We then apply the procedure defined in Lemma~\ref{sdvrp} to the tour in $\{C_\varepsilon, C_\alpha\}$ with the smaller cost to compute a set of tours $\T$ serving all customers, where each tour serves at most $Q$ demand.

Clearly, the algorithm runs in polynomial time.
Moreover, by Lemma~\ref{sdvrp}, $\T$ utilizes $\ceil{\frac{\sum_{v\in J}q_v}{Q}}$ vehicles. 
Since $\sum_{v\in J}q_v\leq mQ$, the resulting solution is also feasible for the SDVRP.
Regarding the routing cost, Lemma~\ref{sdvrp} implies that $c(\T)\leq \min\{c(C_\varepsilon),c(C_\alpha)\} + \frac{\sum_{v\in J}q_v\cdot 2c(u,v)}{Q}$.

If $\frac{\sum_{v\in J}q_v\cdot 2c(u,v)}{Q}\geq (1-\varepsilon)\cdot \OPT'$, then $c(C_\varepsilon)\leq (1+f(\varepsilon))\cdot \OPT'$, and by the VRP lower bounds, $c(\T)\leq c(C_\varepsilon) + \frac{\sum_{v\in J}q_v\cdot 2c(u,v)}{Q}\leq (2+f(\varepsilon))\cdot \OPT'$. Otherwise, $c(\T)\leq c(C_\alpha) + \frac{\sum_{v\in J}q_v\cdot 2c(u,v)}{Q}\leq (\alpha+1-\varepsilon)\cdot \OPT'$.
Thus, by selecting the constant $\varepsilon = \varepsilon_\alpha$ such that $2+f(\varepsilon_\alpha)=\alpha+1-\varepsilon_\alpha$, the algorithm achieves a ratio of $\alpha+1-\varepsilon_\alpha$, matching the ratio for the splittable VRP~\citep{journals/mp/BlauthTV23}.
\end{proof}

\section{Experiments}\label{sec8}
The performance of \textsc{Alg.1} depends on the parameter $\varepsilon$. 
For practical purposes and simplicity, we propose a simplified FPT variant, denoted by \textsc{Alg.1+}, based on the structural insights of \textsc{Alg.1}.

\textsc{Alg.1+} first computes a minimum-cost spanning forest $\F$ such that each tree contains exactly one depot~\citep{journals/ijoc/XuR15}. It then enumerates all possible partitions $\mathfrak{F}$ of the trees in $\F$ with $\sum_{\F_i\in\mathfrak{F}}\Ceil{\frac{\sum_{v\in J(\F_i)}q_v}{Q}}\leq m$ as in line~\ref{alg:main-partition} of \textsc{Alg.1}. 
For each $\F_i \in \mathfrak{F}$, the algorithm connects all trees in $\F_i$ into a single component by repeatedly adding a minimum-cost edge that connects two vertex-disjoint trees, similar to Kruskal's algorithm.
Let $E_i$ denote the set of added edges such that $C_i = (V(\F_i), E_i \cup E(\F_i))$ forms a component, and let $\S' = \{C_i\}_{\F_i \in \mathfrak{F}}$ denote the resulting component set. 
Next, \textsc{Alg.1+} computes a minimum-cost perfect matching $M$ on the odd-degree vertices of $\S'$ so that $M \cup E(\S')$ forms an even-degree graph.
Finally, \textsc{Alg.1+} shortcuts this graph into a cycle cover $\S$ and obtains a solution $\T$ by calling \textsc{Transform}, as in line~\ref{alg:main-transform} of \textsc{Alg.1}. In Appendix~\ref{appB}, we prove that \textsc{Alg.1+} runs in $2^{\O(k\log k)}\cdot \O(\size{V}^4\log\size{V})$ time and achieves an approximation ratio of $9$.

We compare our \textsc{Alg.1+} and \textsc{Alg.2} against the previous XP $(6-4/k)$-approximation algorithm (\textsc{Appx}) and the baseline heuristic (\textsc{Greedy}) from \citet{journals/ijoc/LaiXXD23}. Note that \textsc{Greedy} is adapted from a classical greedy heuristic for VRPs~\citep{journals/jors/MoleJ76}.

\paragraph{Implementation.} 
Both \textsc{Alg.1+} and \textsc{Appx} require computing a minimum-weight perfect matching; for this purpose, we use \textsc{Blossom V}~\citep{journals/mpc/Kolmogorov09} to improve computational efficiency. 
To solve the MCMFP, rather than employing the \emph{strongly} polynomial-time algorithm~\citep{DBLP:journals/ior/Orlin93}, we use the cost-scaling algorithm~\citep{journals/moor/GoldbergT90} provided by the LEMON C++ graph library~\citep{journals/entcs/DezsoJK11}. 
It executes highly efficiently, despite having a \emph{weakly} polynomial running time of $\O(\size{V}^2 \size{E}\log(\size{V}\cdot c_{\max}))$, where $c_{\max}$ is the maximum edge cost.
Moreover, we impose a runtime limit of 3600 seconds for all algorithms in our experiments. 

All algorithms are implemented in C++, compiled with \texttt{g++ -O3}, and executed on a server with an Intel Xeon Gold 6226R CPU @ 2.90GHz, 512 GB RAM, running Ubuntu Server 20.04 LTS. The source code is available at \url{https://github.com/Serryh1/MD-SDVRP}.
 
\paragraph{Instances.} 
We initially evaluate the algorithms on two datasets (P Set and SD Set) from \citet{journals/ijoc/LaiXXD23}, which were generated by adapting benchmark instances of the MD-VRP and SDVRP. 
The P Set contains 18 instances, and the SD Set contains 63 instances, split evenly with 21 instances (SD1--SD21) for each depot count $k\in\{2,4,6\}$.
All instances satisfy $q_v < Q$. 
Notably, for each $k$, instances SD2--SD21 possess the tight capacity property $mQ-\sum_{v\in J}q_v=0$.

To evaluate performance on larger instances, a higher number of depots, and scenarios where $q_v \geq Q$, we generate an additional dataset of 20 instances (the G Set) using an MD-VRP generation method adapted from \citep{journals/transci/HarksKM13}.
We vary two parameters: the instance size and the vehicle capacity.
The instance size combinations $(\size{J}, k)$ take five values: $(500, 10)$, $(750, 12)$, $(1000, 14)$, $(1250, 16)$, and $(1500, 18)$. The vehicle capacity $Q$ takes four values: 100, 200, 300, and 400. For each instance, the number of vehicles at each depot and the demand of each customer are integers sampled uniformly from $[4,100]$ and $[1,Q+100]$, respectively.
Customers and depots are situated in the Euclidean plane, with coordinates sampled uniformly from $[-1000,1000]$.

\paragraph{Results.} 
We consider two primary evaluation metrics: running time and the normalized solution value~(NS-value), computed as $\frac{\text{Algorithm Solution Value}}{\text{\textsc{Appx} Solution Value}} \times 100\%$.  

\begin{figure*}[t]
    \centering
    \includegraphics[width=\linewidth]{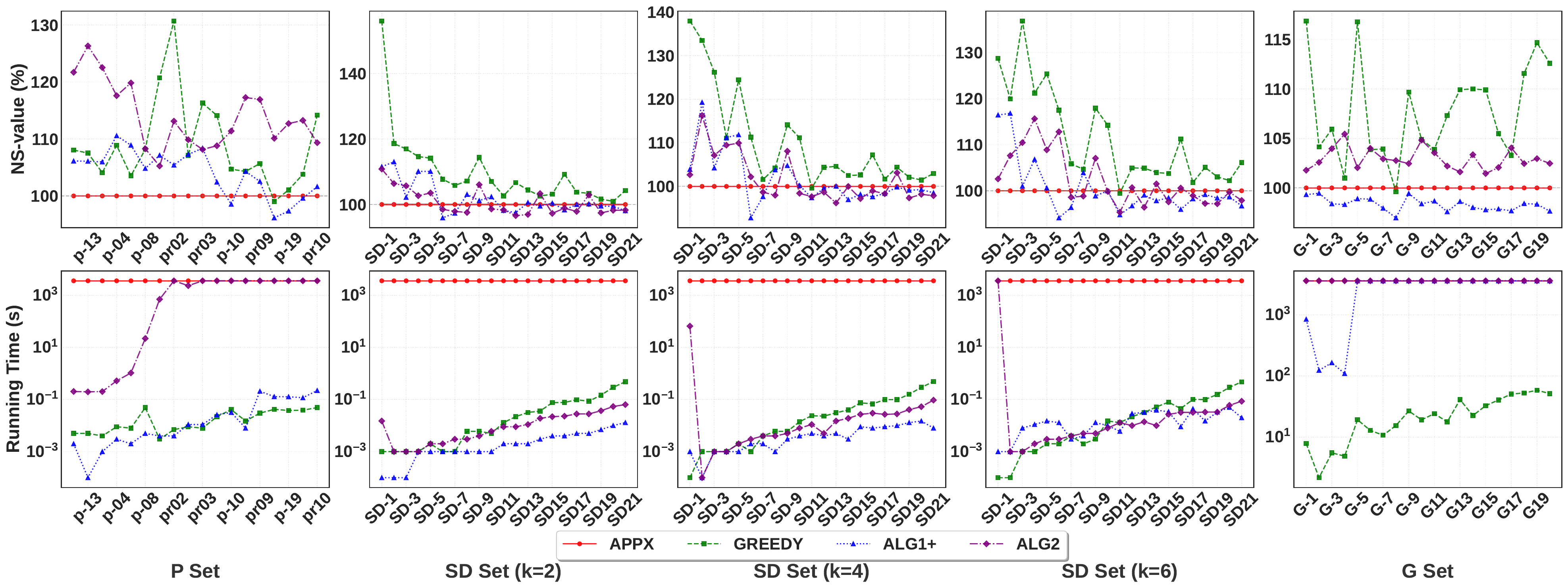}
    \caption{Experimental results across datasets. The top row depicts the NS-value, and the bottom row depicts the running time. Columns correspond to the datasets (from left to right): P Set, SD Set ($k\in \{2, 4, 6 \}$), and G Set. Results for \textsc{Appx} (previous), \textsc{Greedy} (previous), \textsc{Alg.1+} (ours), and \textsc{Alg.2} (ours) are plotted in green, red, blue, and purple, respectively.
    }
    \label{fig1}
\end{figure*}

Figure~\ref{fig1} shows the overall experimental results, with detailed numerical data given in Appendix~\ref{appC}.

On the P Set, \textsc{Appx} demonstrates the best overall performance, being outperformed by \textsc{Alg.1+} on only four instances. Both \textsc{Alg.2} and \textsc{Greedy} underperform relative to \textsc{Appx}. 
Conversely, on the SD Set, \textsc{Alg.1+} (best on 22 of 63 instances) and \textsc{Alg.2} (best on 23 of 63 instances) yield the most competitive results, achieving this with running times significantly faster than \textsc{Appx}. 
For the large-scale G Set, \textsc{Alg.1+} obtains the lowest reported solution cost across all instances under the time limit. \textsc{Appx} and \textsc{Alg.2} return suboptimal solutions, and \textsc{Greedy}, while computationally fast, consistently produces the poorest solution quality.

\paragraph{Summary.} 
Overall, \textsc{Alg.1+} gives competitive solution quality on the tested instances, particularly on the larger G Set. The reported results include runs reaching the time limit, so they should not be interpreted as completion times for the full enumeration.
While \textsc{Appx} performs well on small-$k$ datasets, it suffers from severe computational inefficiency as instance size grows.
\textsc{Alg.2} holds a distinct advantage on the SD Set, especially for instances where $mQ - \sum_{v\in J} q_v = 0$.
Finally, \textsc{Greedy} requires less computation, but its solution costs are higher on the SD and G datasets.

\bibliographystyle{apalike} 
\bibliography{main} 

\begin{thebibliography}{}

\bibitem[Adamaszek et~al., 2018]{DBLP:conf/stacs/AdamaszekA0M18}
Adamaszek, A., Antoniadis, A., Kumar, A., and M{\"{o}}mke, T. (2018).
\newblock Approximating airports and railways.
\newblock In {\em {STACS} 2018}, pages 5:1--5:13.

\bibitem[Adamaszek et~al., 2016]{DBLP:conf/stacs/AdamaszekAM16}
Adamaszek, A., Antoniadis, A., and M{\"{o}}mke, T. (2016).
\newblock Airports and railways: {F}acility location meets network design.
\newblock In {\em {STACS} 2016}, pages 6:1--6:14.

\bibitem[Aggarwal et~al., 2013]{journals/mp/AggarwalLBGGGJ13}
Aggarwal, A., Louis, A., Bansal, M., Garg, N., Gupta, N., Gupta, S., and Jain, S. (2013).
\newblock A $3$-approximation algorithm for the facility location problem with uniform capacities.
\newblock {\em Math. Program.}, 141(1):527--547.

\bibitem[Altinkemer and Gavish, 1987]{journals/orl/AltinkemerG87}
Altinkemer, K. and Gavish, B. (1987).
\newblock Heuristics for unequal weight delivery problems with a fixed error guarantee.
\newblock {\em Oper. Res. Lett.}, 6(4):149--158.

\bibitem[Archetti et~al., 2014]{DBLP:journals/eor/ArchettiBS14a}
Archetti, C., Bianchessi, N., and Speranza, M.~G. (2014).
\newblock Branch-and-cut algorithms for the split delivery vehicle routing problem.
\newblock {\em Eur. J. Oper. Res.}, 238(3):685--698.

\bibitem[Archetti et~al., 2006]{DBLP:journals/transci/ArchettiSS06}
Archetti, C., Savelsbergh, M. W.~P., and Speranza, M.~G. (2006).
\newblock Worst-case analysis for split delivery vehicle routing problems.
\newblock {\em Transp. Sci.}, 40(2):226--234.

\bibitem[Bansal et~al., 2012]{conf/esa/BansalGG12}
Bansal, M., Garg, N., and Gupta, N. (2012).
\newblock A $5$-approximation for capacitated facility location.
\newblock In {\em {ESA} 2012}, pages 133--144.

\bibitem[Belenguer et~al., 2000]{journals/ior/BelenguerMM00}
Belenguer, J.-M., Martinez, M., and Mota, E. (2000).
\newblock A lower bound for the split delivery vehicle routing problem.
\newblock {\em Oper. Res.}, 48(5):801--810.

\bibitem[Blauth et~al., 2023]{journals/mp/BlauthTV23}
Blauth, J., Traub, V., and Vygen, J. (2023).
\newblock Improving the approximation ratio for capacitated vehicle routing.
\newblock {\em Math. Program.}, 197(2):451--497.

\bibitem[Chen et~al., 2025]{DBLP:journals/jacm/ChenKLPGS25}
Chen, L., Kyng, R., Liu, Y.~P., Peng, R., Gutenberg, M.~P., and Sachdeva, S. (2025).
\newblock Maximum flow and minimum-cost flow in almost-linear time.
\newblock {\em J. {ACM}}, 72(3):19:1--19:103.

\bibitem[Chen, 2023]{conf/iccmso/Chen23}
Chen, Y. (2023).
\newblock Approximation schemes for capacity vehicle routing problems: {A} survey.
\newblock In {\em {ICCMSO} 2023}, pages 277--282.

\bibitem[Christofides, 1976]{reports/cmu/Christofides76}
Christofides, N. (1976).
\newblock Worst-case analysis of a new heuristic for the travelling salesman problem.
\newblock Technical report, Carnegie-Mellon University.

\bibitem[Contardo et~al., 2014]{journals/ijoc/ContardoCG14}
Contardo, C., Cordeau, J.-F., and Gendron, B. (2014).
\newblock An exact algorithm based on cut-and-column generation for the capacitated location-routing problem.
\newblock {\em INFORMS J. Comput.}, 26(1):88--102.

\bibitem[Cormen et~al., 2022]{books/mit/CormenLRS22}
Cormen, T.~H., Leiserson, C.~E., Rivest, R.~L., and Stein, C. (2022).
\newblock {\em {I}ntroduction to {A}lgorithms}.
\newblock MIT Press.

\bibitem[Das and Mathieu, 2015]{journals/algorithmica/DasM15}
Das, A. and Mathieu, C. (2015).
\newblock A quasipolynomial time approximation scheme for euclidean capacitated vehicle routing.
\newblock {\em Algorithmica}, 73(1):115--142.

\bibitem[Das et~al., 2020]{DBLP:conf/approx/DasJK20}
Das, S., Jain, L., and Kumar, N. (2020).
\newblock A constant factor approximation for capacitated min-max tree cover.
\newblock In {\em {APPROX/RANDOM} 2020}, pages 55:1--55:13.

\bibitem[Deppert et~al., 2023]{conf/esa/DeppertKM23}
Deppert, M., Kaul, M., and Mnich, M. (2023).
\newblock A $(3/2 + \varepsilon)$-approximation for multiple {TSP} with a variable number of depots.
\newblock In {\em {ESA} 2023}, pages 39:1--39:15.

\bibitem[Dezs{\H{o}} et~al., 2011]{journals/entcs/DezsoJK11}
Dezs{\H{o}}, B., J{\"u}ttner, A., and Kov{\'a}cs, P. (2011).
\newblock {LEMON}--an open source c++ graph template library.
\newblock {\em Electron. Notes Theor. Comput. Sci.}, 264(5):23--45.

\bibitem[Dror and Trudeau, 1989]{DBLP:journals/transci/DrorT89}
Dror, M. and Trudeau, P. (1989).
\newblock Savings by split delivery routing.
\newblock {\em Transp. Sci.}, 23(2):141--145.

\bibitem[Dror and Trudeau, 1990]{journals/nrl/DrorT90}
Dror, M. and Trudeau, P. (1990).
\newblock Split delivery routing.
\newblock {\em Naval Research Logistics}, 37(3):383--402.

\bibitem[Fakcharoenphol et~al., 2004]{DBLP:journals/jcss/FakcharoenpholRT04}
Fakcharoenphol, J., Rao, S., and Talwar, K. (2004).
\newblock A tight bound on approximating arbitrary metrics by tree metrics.
\newblock {\em J. Comput. Syst. Sci.}, 69(3):485--497.

\bibitem[Friggstad et~al., 2026]{conf/soda/FriggstadGM26}
Friggstad, Z., Grandoni, F., and Mousavi, R. (2026).
\newblock Breaching the $2$-approximation barrier for euclidean capacitated vehicle routing.
\newblock In {\em {SODA} 2026}, pages 2644--2668.

\bibitem[Friggstad and M{\"{o}}mke, 2025]{DBLP:journals/corr/abs-2510-05321}
Friggstad, Z. and M{\"{o}}mke, T. (2025).
\newblock Approximating multiple-depot capacitated vehicle routing via {LP} rounding.
\newblock {\em CoRR}, abs/2510.05321.

\bibitem[Friggstad et~al., 2025]{journals/moor/FriggstadMRS25}
Friggstad, Z., Mousavi, R., Rahgoshay, M., and Salavatipour, M.~R. (2025).
\newblock Improved approximations for a {CVRP} with unsplittable demands.
\newblock {\em Math. Oper. Res.}, 0(0).

\bibitem[Gamst et~al., 2024]{journals/transci/GamstLR24}
Gamst, M., Lusby, R.~M., and Ropke, S. (2024).
\newblock Exact and heuristic methods for the split delivery vehicle routing problem.
\newblock {\em Transp. Sci.}, 58(4):741--760.

\bibitem[Goemans and Williamson, 1995]{DBLP:journals/siamcomp/GoemansW95}
Goemans, M.~X. and Williamson, D.~P. (1995).
\newblock A general approximation technique for constrained forest problems.
\newblock {\em {SIAM} J. Comput.}, 24(2):296--317.

\bibitem[Goldberg and Tarjan, 1990]{journals/moor/GoldbergT90}
Goldberg, A.~V. and Tarjan, R.~E. (1990).
\newblock Finding minimum-cost circulations by successive approximation.
\newblock {\em Math. Oper. Res.}, 15(3):430--466.

\bibitem[G{\o}rtz and Nagarajan, 2016]{journals/networks/GortzN16}
G{\o}rtz, I.~L. and Nagarajan, V. (2016).
\newblock Locating depots for capacitated vehicle routing.
\newblock {\em Networks}, 68(2):94--103.

\bibitem[Gouveia et~al., 2023]{journals/transci/GouveiaLR23}
Gouveia, L., Leitner, M., and Ruthmair, M. (2023).
\newblock Multi-depot routing with split deliveries: {M}odels and a branch-and-cut algorithm.
\newblock {\em Transp. Sci.}, 57(2):512--530.

\bibitem[Gutin et~al., 2017]{DBLP:journals/jcss/GutinWY17}
Gutin, G.~Z., Wahlstr{\"{o}}m, M., and Yeo, A. (2017).
\newblock Rural postman parameterized by the number of components of required edges.
\newblock {\em J. Comput. Syst. Sci.}, 83(1):121--131.

\bibitem[Haimovich and Kan, 1985]{journals/moor/HaimovichK85}
Haimovich, M. and Kan, A. H. G.~R. (1985).
\newblock Bounds and heuristics for capacitated routing problems.
\newblock {\em Math. Oper. Res.}, 10(4):527--542.

\bibitem[Harks et~al., 2013]{journals/transci/HarksKM13}
Harks, T., K{\"{o}}nig, F.~G., and Matuschke, J. (2013).
\newblock Approximation algorithms for capacitated location routing.
\newblock {\em Transp. Sci.}, 47(1):3--22.

\bibitem[He et~al., 2025]{journals/ijoc/HeHW25}
He, P., Hao, J.-K., and Wu, Q. (2025).
\newblock A hybrid genetic algorithm with multi-population for capacitated location routing.
\newblock {\em INFORMS J. Comput.}, 0(0).

\bibitem[Heine et~al., 2023]{journals/eor/HeineDM23}
Heine, F.~C., Demleitner, A., and Matuschke, J. (2023).
\newblock Bifactor approximation for location routing with vehicle and facility capacities.
\newblock {\em Eur. J. Oper. Res.}, 304(2):429--442.

\bibitem[Hu, 2009]{phd/sfu/Hu09}
Hu, Y. (2009).
\newblock {\em Approximation algorithms for the capacitated vehicle routing problem}.
\newblock PhD thesis, Simon Fraser University.

\bibitem[Jayaprakash and Salavatipour, 2023]{journals/talg/JayaprakashS23}
Jayaprakash, A. and Salavatipour, M.~R. (2023).
\newblock Approximation schemes for capacitated vehicle routing on graphs of bounded treewidth, bounded doubling, or highway dimension.
\newblock {\em ACM Trans. Algorithms}, 19(2):1--36.

\bibitem[Jowhari and Nematollahi, 2025]{DBLP:journals/ipl/JowhariN25}
Jowhari, H. and Nematollahi, S. (2025).
\newblock Airports and railways with unsplittable demand.
\newblock {\em Inf. Process. Lett.}, 188:106538.

\bibitem[Karlin et~al., 2023]{DBLP:conf/ipco/KarlinKG23}
Karlin, A.~R., Klein, N., and Gharan, S.~O. (2023).
\newblock A deterministic better-than-$3/2$ approximation algorithm for metric {TSP}.
\newblock In {\em {IPCO} 2023}, pages 261--274.

\bibitem[Karlin et~al., 2021]{DBLP:conf/stoc/KarlinKG21}
Karlin, A.~R., Klein, N., and Oveis~Gharan, S. (2021).
\newblock {A} (slightly) improved approximation algorithm for metric {TSP}.
\newblock In {\em {STOC} 2021}, pages 32--45.

\bibitem[Kolmogorov, 2009]{journals/mpc/Kolmogorov09}
Kolmogorov, V. (2009).
\newblock Blossom {V}: {A} new implementation of a minimum cost perfect matching algorithm.
\newblock {\em Math. Program. Comput.}, 1(1):43--67.

\bibitem[Konstantakopoulos et~al., 2022]{journals/or/KonstantakopoulosGK22}
Konstantakopoulos, G.~D., Gayialis, S.~P., and Kechagias, E.~P. (2022).
\newblock Vehicle routing problem and related algorithms for logistics distribution: A literature review and classification.
\newblock {\em Oper. Res.}, 22(3):2033--2062.

\bibitem[Lai et~al., 2023]{journals/ijoc/LaiXXD23}
Lai, X., Xu, L., Xu, Z., and Du, Y. (2023).
\newblock An approximation algorithm for $k$-depot split delivery vehicle routing problem.
\newblock {\em INFORMS J. Comput.}, 35(5):1179--1194.

\bibitem[Li and Simchi{-}Levi, 1990]{journals/ijoc/LiS90}
Li, C. and Simchi{-}Levi, D. (1990).
\newblock Worst-case analysis of heuristics for multidepot capacitated vehicle routing problems.
\newblock {\em INFORMS J. Comput.}, 2(1):64--73.

\bibitem[Lin et~al., 2025]{journals/ijoc/LinHJMSL25}
Lin, W., He, Z., Jiang, S., Ma, F., Su, Z., and L{\"u}, Z. (2025).
\newblock Alkaid-sdvrp: {A}n efficient open-source solver for the vehicle routing problem with split deliveries.
\newblock {\em INFORMS J. Comput.}, 0(0).

\bibitem[Mathieu and Zhou, 2023]{journals/talg/MathieuZ23}
Mathieu, C. and Zhou, H. (2023).
\newblock A {PTAS} for capacitated vehicle routing on trees.
\newblock {\em ACM Trans. Algorithms}, 19(2):1--28.

\bibitem[Mathieu and Zhou, 2025]{journals/mp/MathieuZ25}
Mathieu, C. and Zhou, H. (2025).
\newblock A tight $(1.5+\varepsilon)$-approximation for unsplittable capacitated vehicle routing on trees.
\newblock {\em Math. Program.}, 212(1):115--146.

\bibitem[Mole and Jameson, 1976]{journals/jors/MoleJ76}
Mole, R. and Jameson, S. (1976).
\newblock A sequential route-building algorithm employing a generalised savings criterion.
\newblock {\em J. Oper. Res. Soc.}, 27(2):503--511.

\bibitem[Montoya-Torres et~al., 2015]{journals/cie/MontoyaTorresFIJH15}
Montoya-Torres, J.~R., Franco, J.~L., Isaza, S.~N., Jim{\'e}nez, H.~F., and Herazo-Padilla, N. (2015).
\newblock A literature review on the vehicle routing problem with multiple depots.
\newblock {\em Comput. Ind. Eng.}, 79:115--129.

\bibitem[Orlin, 1993]{DBLP:journals/ior/Orlin93}
Orlin, J.~B. (1993).
\newblock A faster strongly polynomial minimum cost flow algorithm.
\newblock {\em Oper. Res.}, 41(2):338--350.

\bibitem[Rathinam et~al., 2007]{journals/tase/RathinamSD07}
Rathinam, S., Sengupta, R., and Darbha, S. (2007).
\newblock A resource allocation algorithm for multivehicle systems with nonholonomic constraints.
\newblock {\em IEEE Trans Autom. Sci. Eng.}, 4(1):98--104.

\bibitem[Salavatipour and Tian, 2025]{DBLP:journals/jco/SalavatipourT25}
Salavatipour, M.~R. and Tian, L. (2025).
\newblock Approximation algorithms for the airport and railway problem.
\newblock {\em J. Comb. Optim.}, 49(1):8.

\bibitem[Schneider and Drexl, 2017]{journals/anor/SchneiderD17}
Schneider, M. and Drexl, M. (2017).
\newblock A survey of the standard location-routing problem.
\newblock {\em Ann. Oper. Res.}, 259(1):389--414.

\bibitem[Schrijver, 2003]{books/springer/Schrijver03}
Schrijver, A. (2003).
\newblock {\em {C}ombinatorial {O}ptimization: {P}olyhedra and {E}fficiency}, volume~24.
\newblock Springer.

\bibitem[Serdyukov, 1978]{journals/us/Serdyukov78}
Serdyukov, A.~I. (1978).
\newblock Some extremal bypasses in graphs.
\newblock {\em Upravlyaemye Sistemy}, 17:76--79.

\bibitem[Toth and Vigo, 2014]{books/siam/TothV14}
Toth, P. and Vigo, D. (2014).
\newblock {\em {V}ehicle {R}outing: {P}roblems, {M}ethods, and {A}pplications}.
\newblock SIAM.

\bibitem[Traub et~al., 2022]{DBLP:journals/siamcomp/TraubVZ22}
Traub, V., Vygen, J., and Zenklusen, R. (2022).
\newblock Reducing path {TSP} to {TSP}.
\newblock {\em {SIAM} J. Comput.}, 51(3):20--24.

\bibitem[Xu and Rodrigues, 2015]{journals/ijoc/XuR15}
Xu, Z. and Rodrigues, B. (2015).
\newblock A $3/2$-approximation algorithm for the multiple {TSP} with a fixed number of depots.
\newblock {\em INFORMS J. Comput.}, 27(4):636--645.

\bibitem[Xu and Rodrigues, 2017]{DBLP:journals/eor/XuR17}
Xu, Z. and Rodrigues, B. (2017).
\newblock An extension of the christofides heuristic for the generalized multiple depot multiple traveling salesmen problem.
\newblock {\em Eur. J. Oper. Res.}, 257(3):735--745.

\bibitem[Xu et~al., 2011]{journals/orl/XuXR11}
Xu, Z., Xu, L., and Rodrigues, B. (2011).
\newblock An analysis of the extended christofides heuristic for the $k$-depot tsp.
\newblock {\em Oper. Res. Lett.}, 39(3):218--223.

\bibitem[Zhao and Xiao, 2025]{DBLP:journals/tcs/ZhaoX25a}
Zhao, J. and Xiao, M. (2025).
\newblock Multidepot capacitated vehicle routing with improved approximation guarantees.
\newblock {\em Theor. Comput. Sci.}, 1043:115265.

\bibitem[Zhao and Xiao, 2026]{DBLP:journals/corr/abs-2601-21660}
Zhao, J. and Xiao, M. (2026).
\newblock Improved approximations for the unsplittable capacitated vehicle routing problem.
\newblock {\em CoRR}, abs/2601.21660.

\bibitem[Zhao et~al., 2024]{DBLP:conf/ijcai/00010W24}
Zhao, J., Xiao, M., and Wang, S. (2024).
\newblock Improved approximation algorithms for capacitated location routing.
\newblock In {\em {IJCAI} 2024}, pages 6805--6813.

\end{thebibliography}

\appendix
\section{A Simplified FPT Approximation Algorithm Parameterized by the Number of Depots}\label{appB}
The details of the simplified algorithm, denoted as \textsc{Alg.1+}, are provided in Algorithm~\ref{alge}.

\begin{algorithm}[!t]
\small
\caption{\textsc{Alg.1+}: An FPT $9$-approximation algorithm for the MD-SDVRP}
\label{alge}
\textbf{Input:} An MD-SDVRP instance $I=(G,c,q,r,Q)$. \\
\textbf{Output:} A feasible solution $\T_f$.
\begin{algorithmic}[1]
\State Initialize $\mathfrak{T}\coloneq \emptyset$.\label{alges1}

\State Find a minimum-cost spanning forest $\F$ in $G$ such that each tree contains only one depot~\citep{journals/ijoc/XuR15}.\label{alges3}

\For{each partition $\mathfrak{F}$ of $\F$ with $\sum_{\F_i\in\mathfrak{F}}\Ceil{\frac{\sum_{v\in J(\F_i)}q_v}{Q}}\leq m$}\label{alges4}
\State For each $\F_i\in\mathfrak{F}$, connect all trees in $\F_i$ by repeatedly adding a minimum-cost edge that connects two vertex-disjoint trees (as in Kruskal's algorithm), and let $E_i$ denote the set of added edges.

\State Obtain a set of components $\S'=\{C_i\}_{\F_i\in\mathfrak{F}}$ with $E(C_i)=E_i\cup E(\F_i)$.
Then, let $O$ denote the set of odd-degree vertices in the component set $\S'$, compute a minimum-cost perfect matching $M$ in $G[O]$ to obtain an even-degree graph with edge set $E(\S')\cup M$, and finally shortcut the even-degree graph to obtain a cycle cover $\S$ in $G$.\label{alges7}

\State Call \textsc{Transform} on $\S$ to obtain a solution $\T$ to $I$, and update $\mathfrak{T}\coloneq\mathfrak{T}\cup\{\T\}$.\label{alges8}
\EndFor\label{alges14}
\State \Return $\T_f=\arg\min_{\T\in\mathfrak{T}}c(\T)$.
\end{algorithmic}
\end{algorithm}

\begin{theorem}\label{res2+}
For the MD-SDVRP, \textsc{Alg.1+} runs in $2^{\O(k\log k)}\cdot \O(\size{V}^4\log\size{V})$ time and achieves an approximation ratio of $9$.
\end{theorem}
\begin{proof}
We analyze the solution quality and the running time of \textsc{Alg.1+}.

First, the forest $\F$ in line~\ref{alges3} can be computed in $\O(\size{V}^2)$ time~\citep{journals/ijoc/XuR15}, and clearly $c(\F)\leq \OPT$.

Second, we show that there exists an edge set $E_J$ from $G[J]$ that augments $\F$ into a good component set and satisfies $c(E_J)\leq\OPT$.
By shortcutting each component in $\S^*$ with at least one customer, we obtain a cycle cover $\B^*=\{C_1,\dots,C_t\}$ in $G[J]$. Since every component of $\S^*$ contains at least one depot, we have $t\leq k$. Moreover, by the triangle inequality, $c(\B^*)\leq\OPT$.

We iteratively construct $E_J$ by processing the cycles in $\B^*$ to merge the components of $\F$.
Let $\S_0=\F$ and initialize $E_J=\emptyset$.
For each $i\in\{1,\dots,t\}$, let $\S'_{i-1}=\{C\in\S_{i-1}\mid V(C)\cap V(C_i)\neq\emptyset\}$ and $m_i=\size{\S'_{i-1}}$.
If $m_i>1$, construct a cycle $C'_i$ by shortcutting $C_i$ so that it visits exactly one vertex in each component in $\S'_{i-1}$, and add its edges to $E_J$. Otherwise, let $C'_i$ be a trivial cycle.
Let $\S_i$ be the set of connected components formed by $E(\F)\cup E_J$.

By construction, all vertices in $V(C_i)$ belong to a single component in $\S_i$, and consequently in $\S_t$.
Since $\B^*$ is obtained from $\S^*$, the component set $\S_t$ is good.
Moreover, by the triangle inequality,
\[
c(E_J)\leq\sum_{i=1}^t c(C'_i)\leq\sum_{i=1}^t c(C_i)\leq\OPT.
\]

Hence, there exists a partition $\mathfrak{F}$ of $\F$ such that for each $\F_i\in\mathfrak{F}$, the edges in $E_J$ connect all trees in $\F_i$ into a single connected component.
Since this partition is good, it satisfies the vehicle-count test by Property~\ref{good}. When it is considered in line~\ref{alges4}, by the greedy nature of Kruskal's algorithm, $E_i$ is a minimum-cost edge set connecting all trees in $\F_i$.
Therefore, we are guaranteed that $\sum_{\F_i\in\mathfrak{F}}c(E_i)\leq c(E_J) \leq \OPT$ in this specific iteration.
Moreover, the component set $\S'$ obtained in line~\ref{alges7} is a good component set, and its cost is bounded by $c(\S')\leq c(\F)+c(E_J)\leq 2\cdot \OPT$.

Next, we bound the cost of the edge set $M$ obtained in line~\ref{alges7}. Since $M$ is a minimum-cost perfect matching on the odd-degree vertices of $\S'$, and one can always augment $\S'$ into an even-degree graph simply by doubling all edges in $E(\S')$, it follows that $c(M)\leq c(\S')$.
Since the cycle cover $\S$ in line~\ref{alges7} is obtained by shortcutting the even-degree graph formed by $E(\S')\cup M$, by the triangle inequality, we have $c(\S)\leq c(\S')+c(M)\leq 2c(\S') \leq 4\cdot \OPT$.

Then, by calling \textsc{Transform} in line~\ref{alges8}, we obtain a solution $\T$ with cost bounded by $c(\T)\leq 2c(\S)+\OPT$ according to Lemma~\ref{flowcost}. Substituting our bound for $c(\S)$, we obtain $c(\T) \leq 2(4\cdot \OPT)+\OPT = 9\cdot \OPT$.

Finally, we analyze the running time. The computation of the spanning forest $\F$ takes $\O(\size{V}^2)$ time. Since each tree in $\F$ contains exactly one depot, $\size{\F} = k$. The number of possible partitions $\mathfrak{F}$ of $\F$ is bounded by the Bell number $B_k \leq k^k = 2^{\O(k\log k)}$. 
For each partition, computing Kruskal's connections takes $\O(\size{V}^2)$ time, the minimum-cost perfect matching takes $\O(\size{V}^3)$ time, and calling \textsc{Transform} takes $\O(\size{V}^4\log\size{V})$ time (by Lemma~\ref{flowtime}).
Therefore, the overall running time of \textsc{Alg.1+} is $2^{\O(k\log k)}\cdot \O(\size{V}^4\log\size{V})$.
\end{proof}


\section{Additional Experimental Results}\label{appC}
The detailed experimental results can be found in Tables~\ref{tab:p_set}--\ref{tab:g_set}.

\begin{table*}[htbp]
    \centering
    \caption{Comparison of \textsc{Appx} with \textsc{Greedy}, \textsc{Alg.1+}, and \textsc{Alg.2} on P Set}
    \resizebox{0.85\linewidth}{!}{ 
    \begin{tabular}{lcccccccccccc}
        \toprule
         & & & & & \multicolumn{2}{c}{\textsc{Appx}} & \multicolumn{2}{c}{\textsc{Greedy}} & \multicolumn{2}{c}{\textsc{Alg.1+}} & \multicolumn{2}{c}{\textsc{Alg.2}}\\
        \cmidrule(lr){6-7} \cmidrule(lr){8-9} \cmidrule(lr){10-11} \cmidrule(lr){12-13}
        Instance & \# Dep. & \# Cus. & \# Veh. &Cap. & OBJ & Time (s) & NS-value (\%) & Time (s) &  NS-value (\%) & Time (s) &  NS-value (\%) & Time (s)\\
        \midrule
        P-12 & 2 & 80 & 10 & 60 & 1,486.47 & 3600.0 & 108.05 & 0.005 & 106.12 & 0.002 & 121.69 & 0.204 \\
        P-13 & 2 & 80 & 10 & 58 & 1,498.47 & 3600.0 & 107.52 & 0.005 & 106.08 & 0.000 & 126.32 & 0.197 \\
        P-14 & 2 & 80 & 10 & 59 & 1,532.82 & 3600.0 & 104.09 & 0.004 & 105.96 & 0.001 & 122.54 & 0.202 \\
        P-04 & 2 & 100 & 16 & 100 & 1,190.47 & 3600.0 & 108.88 & 0.009 & 110.56 & 0.003 & 117.59 & 0.521 \\
        P-05 & 2 & 100 & 10 & 200 & 923.64 & 3600.0 & 103.54 & 0.008 & 108.90 & 0.002 & 119.83 & 1.050 \\
        P-08 & 2 & 249 & 28 & 500 & 5,259.69 & 3600.0 & 108.19 & 0.049 & 104.83 & 0.005 & 108.27 & 21.987 \\
        P-01 & 4 & 50 & 16 & 80 & 697.48 & 3600.0 & 120.74 & 0.003 & 107.14 & 0.004 & 105.26 & 703.053 \\
        Pr02 & 4 & 96 & 8 & 195 & 1,498.91 & 3600.0 & 130.71 & 0.007 & 105.41 & 0.004 & 113.13 & 3600.000 \\
        P-07 & 4 & 100 & 16 & 100 & 1,060.26 & 3600.0 & 107.12 & 0.009 & 107.28 & 0.011 & 109.86 & 2344.804 \\
        Pr03 & 4 & 144 & 12 & 190 & 2,195.75 & 3600.0 & 116.29 & 0.008 & 108.37 & 0.011 & 108.16 & 3600.000 \\
        Pr04 & 4 & 192 & 16 & 185 & 2,469.53 & 3600.0 & 114.11 & 0.022 & 102.43 & 0.026 & 108.81 & 3600.000 \\
        P-10 & 4 & 249 & 32 & 500 & 4,675.89 & 3600.0 & 104.69 & 0.041 & \textbf{98.54} & 0.032 & 111.39 & 3600.000 \\
        Pr08 & 6 & 144 & 12 & 190 & 2,067.40 & 3600.0 & 104.30 & 0.015 & 104.30 & 0.008 & 117.27 & 3600.000 \\
        Pr09 & 6 & 216 & 18 & 180 & 2,702.20 & 3600.0 & 105.69 & 0.030 & 102.54 & 0.208 & 116.93 & 3600.000 \\
        P-18 & 6 & 240 & 30 & 60 & 5,155.18 & 3600.0 & 99.03 & 0.042 & \textbf{96.16} & 0.129 & 110.11 & 3600.000 \\
        P-19 & 6 & 240 & 30 & 59 & 5,142.44 & 3600.0 & 101.05 & 0.038 & \textbf{97.35} & 0.128 & 112.70 & 3600.000 \\
        P-20 & 6 & 240 & 30 & 58 & 5,139.05 & 3600.0 & 103.79 & 0.039 & \textbf{99.59} & 0.117 & 113.26 & 3600.000 \\
        Pr10 & 6 & 288 & 24 & 170 & 3,526.01 & 3600.0 & 114.19 & 0.049 & 101.63 & 0.222 & 109.34 & 3600.000 \\
        \midrule
        Average & 4 & 160 & 14 & 163.5 & 2,678.98 & 3600.0 & 109.00 & 0.022 & 104.07 & 0.051 & 114.03 & 2173.543 \\
        \bottomrule
    \end{tabular}
    }
    \label{tab:p_set}
\end{table*}

\begin{table*}[htbp]
    \centering
    \caption{Comparison of \textsc{Appx} with \textsc{Greedy}, \textsc{Alg.1+}, and \textsc{Alg.2} on SD Set ($k=2$)}
    \resizebox{0.85\linewidth}{!}{ 
    \begin{tabular}{lcccccccccccc}
        \toprule
         & & & & & \multicolumn{2}{c}{\textsc{Appx}} & \multicolumn{2}{c}{\textsc{Greedy}} & \multicolumn{2}{c}{\textsc{Alg.1+}} & \multicolumn{2}{c}{\textsc{Alg.2}}\\
        \cmidrule(lr){6-7} \cmidrule(lr){8-9} \cmidrule(lr){10-11} \cmidrule(lr){12-13}
        Instance & \# Dep. & \# Cus. & \# Veh. & Cap. & OBJ & Time (s) & NS-value (\%) & Time (s) &  NS-value (\%) & Time (s) &  NS-value (\%) & Time (s)\\
        \midrule
        SD-1 & 2 & 8 & 8 & 100 & 24,485.28 & 3600.0 & 156.07 & 0.001 & 111.55 & 0.000 & 110.82 & 0.015 \\
        SD-2 & 2 & 16 & 12 & 100 & 70,867.92 & 3600.0 & 118.60 & 0.001 & 113.01 & 0.000 & 106.38 & 0.001 \\
        SD-3 & 2 & 16 & 12 & 100 & 46,954.38 & 3600.0 & 116.97 & 0.001 & 102.17 & 0.000 & 105.70 & 0.001 \\
        SD-4 & 2 & 24 & 18 & 100 & 64,713.66 & 3600.0 & 114.63 & 0.001 & 110.11 & 0.001 & 102.67 & 0.001 \\
        SD-5 & 2 & 32 & 24 & 100 & 138,386.97 & 3600.0 & 114.16 & 0.002 & 110.10 & 0.001 & 103.52 & 0.002 \\
        SD-6 & 2 & 32 & 24 & 100 & 82,799.76 & 3600.0 & 107.73 & 0.001 & \textbf{95.92} & 0.001 & 98.51 & 0.002 \\
        SD-7 & 2 & 40 & 30 & 100 & 349,649.90 & 3600.0 & 105.81 & 0.001 & \textbf{97.32} & 0.001 & 97.89 & 0.003 \\
        SD-8 & 2 & 48 & 36 & 100 & 497,181.08 & 3600.0 & 107.17 & 0.006 & 103.08 & 0.001 & \textbf{97.55} & 0.003 \\
        SD-9 & 2 & 48 & 36 & 100 & 202,809.47 & 3600.0 & 114.38 & 0.006 & 101.19 & 0.001 & 106.06 & 0.004 \\
        SD10 & 2 & 64 & 48 & 100 & 265,479.79 & 3600.0 & 107.04 & 0.005 & 102.36 & 0.001 & \textbf{98.61} & 0.006 \\
        SD11 & 2 & 80 & 60 & 100 & 1,299,085.93 & 3600.0 & 102.63 & 0.013 & \textbf{98.17} & 0.002 & 98.33 & 0.009 \\
        SD12 & 2 & 80 & 60 & 100 & 714,401.10 & 3600.0 & 106.69 & 0.022 & 97.50 & 0.002 & \textbf{96.59} & 0.009 \\
        SD13 & 2 & 96 & 72 & 100 & 1,001,735.52 & 3600.0 & 104.42 & 0.032 & 100.60 & 0.002 & \textbf{96.94} & 0.011 \\
        SD14 & 2 & 120 & 90 & 100 & 1,058,931.41 & 3600.0 & 102.62 & 0.036 & \textbf{99.49} & 0.003 & 103.34 & 0.019 \\
        SD15 & 2 & 144 & 108 & 100 & 1,483,832.65 & 3600.0 & 103.13 & 0.076 & 100.41 & 0.004 & \textbf{97.28} & 0.022 \\
        SD16 & 2 & 144 & 108 & 100 & 312,664.89 & 3600.0 & 109.25 & 0.077 & \textbf{98.36} & 0.004 & 98.87 & 0.023 \\
        SD17 & 2 & 160 & 120 & 100 & 2,625,304.76 & 3600.0 & 103.76 & 0.098 & 99.97 & 0.005 & \textbf{97.87} & 0.028 \\
        SD18 & 2 & 160 & 120 & 100 & 1,395,751.02 & 3600.0 & 103.40 & 0.086 & 100.18 & 0.005 & 102.89 & 0.028 \\
        SD19 & 2 & 192 & 144 & 100 & 1,969,128.17 & 3600.0 & 101.63 & 0.147 & 99.49 & 0.007 & \textbf{97.42} & 0.037 \\
        SD20 & 2 & 240 & 180 & 100 & 3,913,636.93 & 3600.0 & 100.96 & 0.295 & 99.46 & 0.010 & \textbf{98.23} & 0.054 \\
        SD21 & 2 & 288 & 312 & 100 & 1,036,606.65 & 3600.0 & 104.25 & 0.481 & \textbf{98.08} & 0.013 & 98.29 & 0.064 \\
        \midrule
        Average & 2 & 107 & 86 & 100 & 878,842.25 & 3600.0 & 109.78 & 0.067 & 101.83 & 0.004 & 100.66 & 0.020 \\
        \bottomrule
    \end{tabular}
    }
    \label{tab:sd_set_k2}
\end{table*}

\begin{table*}[htbp]
    \centering
    \caption{Comparison of \textsc{Appx} with \textsc{Greedy}, \textsc{Alg.1+}, and \textsc{Alg.2} on SD Set ($k=4$)}
    \resizebox{0.85\linewidth}{!}{ 
    \begin{tabular}{lcccccccccccc}
        \toprule
         & & & & & \multicolumn{2}{c}{\textsc{Appx}} & \multicolumn{2}{c}{\textsc{Greedy}} & \multicolumn{2}{c}{\textsc{Alg.1+}} & \multicolumn{2}{c}{\textsc{Alg.2}}\\
        \cmidrule(lr){6-7} \cmidrule(lr){8-9} \cmidrule(lr){10-11} \cmidrule(lr){12-13}
        Instance & \# Dep. & \# Cus. & \# Veh. &Cap. & OBJ & Time (s) & NS-value (\%) & Time (s) &  NS-value (\%) & Time (s) &  NS-value (\%) & Time (s)\\
        \midrule
        SD-1 & 4 & 8 & 8 & 100 & 19,543.20 & 3600.0 & 137.94 & 0.000 & 103.91 & 0.001 & 102.67 & 65.745 \\
        SD-2 & 4 & 16 & 12 & 100 &  60,903.39 & 3600.0 & 133.48 & 0.001 & 119.27 & 0.000 & 116.26 & 0.000 \\
        SD-3 & 4 & 16 & 12 & 100 & 37,251.43 & 3600.0 & 126.19 & 0.001 & 104.23 & 0.001 & 107.13 & 0.001 \\
        SD-4 & 4 & 24 & 18 & 100 & 51,532.83 & 3600.0 & 111.05 & 0.001 & 111.21 & 0.001 & 109.45 & 0.001 \\
        SD-5 & 4 & 32 & 24 & 100 & 119,754.34 & 3600.0 & 124.44 & 0.002 & 111.84 & 0.001 & 109.93 & 0.002 \\
        SD-6 & 4 & 32 & 24 & 100 & 66,562.03 & 3600.0 & 111.29 & 0.001 & \textbf{92.79} & 0.002 & 102.19 & 0.003 \\
        SD-7 & 4 & 40 & 30 & 100 & 334,352.43 & 3600.0 & 101.57 & 0.004 & \textbf{97.65} & 0.002 & 98.67 & 0.004 \\
        SD-8 & 4 & 48 & 36 & 100 & 478,311.34 & 3600.0 & 104.21 & 0.006 & 103.79 & 0.001 & \textbf{97.97} & 0.004 \\
        SD-9 & 4 & 48 & 36 & 100 & 177,204.03 & 3600.0 & 114.10 & 0.006 & 104.76 & 0.003 & 108.07 & 0.005 \\
        SD10 & 4 & 64 & 48 & 100 & 236,170.62 & 3600.0 & 111.14 & 0.014 & 100.20 & 0.004 & \textbf{98.41} & 0.008 \\
        SD11 & 4 & 80 & 60 & 100 & 1,285,653.88 & 3600.0 & 99.58 & 0.024 & \textbf{97.39} & 0.005 & 97.66 & 0.011 \\
        SD12 & 4 & 80 & 60 & 100 & 674,735.88 & 3600.0 & 104.36 & 0.023 & 99.58 & 0.004 & \textbf{98.67} & 0.005 \\
        SD13 & 4 & 96 & 72 & 100 & 958,943.12 & 3600.0 & 104.60 & 0.031 & 100.05 & 0.005 & \textbf{96.19} & 0.015 \\
        SD14 & 4 & 120 & 90 & 100 & 1,000,936.00 & 3600.0 & 102.49 & 0.040 & \textbf{96.95} & 0.003 & 99.99 & 0.019 \\
        SD15 & 4 & 144 & 108 & 100 & 1,420,590.07 & 3600.0 & 102.63 & 0.075 & 98.18 & 0.009 & \textbf{97.20} & 0.027 \\
        SD16 & 4 & 144 & 108 & 100 & 236,284.80 & 3600.0 & 107.21 & 0.068 & \textbf{97.63} & 0.008 & 98.97 & 0.030 \\
        SD17 & 4 & 160 & 120 & 100 & 2,550,594.21 & 3600.0 & 101.68 & 0.099 & 98.38 & 0.009 & \textbf{98.31} & 0.027 \\
        SD18 & 4 & 160 & 120 & 100 & 1,322,047.51 & 3600.0 & 104.40 & 0.099 & \textbf{99.84} & 0.010 & 103.13 & 0.028 \\
        SD19 & 4 & 192 & 144 & 100 & 1,881,830.25 & 3600.0 & 102.09 & 0.159 & 99.11 & 0.013 & \textbf{97.34} & 0.041 \\
        SD20 & 4 & 240 & 180 & 100 & 3,808,330.88 & 3600.0 & 101.47 & 0.293 & 99.27 & 0.015 & \textbf{98.17} & 0.053 \\
        SD21 & 4 & 288 & 312 & 100 & 892,041.08 & 3600.0 & 102.97 & 0.487 & 98.53 & 0.008 & \textbf{97.87} & 0.095 \\
        \midrule
        Average & 4 & 107 & 86 & 100 & 838,741.59 & 3600.0 & 109.95 & 0.068 & 101.65 & 0.005 & 101.63 & 3.149 \\
        \bottomrule
    \end{tabular}
    }
    \label{tab:sd_set_k4}
\end{table*}

\begin{table*}[htbp]
    \centering
    \caption{Comparison of \textsc{Appx} with \textsc{Greedy}, \textsc{Alg.1+}, and \textsc{Alg.2} on SD Set ($k=6$)}
    \resizebox{0.85\linewidth}{!}{ 
    \begin{tabular}{lcccccccccccc}
        \toprule
         & & & & & \multicolumn{2}{c}{\textsc{Appx}} & \multicolumn{2}{c}{\textsc{Greedy}} & \multicolumn{2}{c}{\textsc{Alg.1+}} & \multicolumn{2}{c}{\textsc{Alg.2}}\\
        \cmidrule(lr){6-7} \cmidrule(lr){8-9} \cmidrule(lr){10-11} \cmidrule(lr){12-13}
        Instance & \# Dep. & \# Cus. & \# Veh. &Cap. & OBJ & Time (s) & NS-value (\%) & Time (s) &  NS-value (\%) & Time (s) &  NS-value (\%) & Time (s)\\
        \midrule
        SD-1 & 6 & 8 & 8 & 100 & 20,307.14 & 3600.0 & 128.73 & 0.000 & 116.45 & 0.001 & 102.57 & 3600.000 \\
        SD-2 & 6 & 16 & 12 & 100 & 61,631.48 & 3600.0 & 119.96 & 0.000 & 116.82 & 0.001 & 107.63 & 0.001 \\
        SD-3 & 6 & 16 & 12 & 100 & 33,069.47 & 3600.0 & 136.88 & 0.001 & 101.00 & 0.008 & 110.46 & 0.001 \\
        SD-4 & 6 & 24 & 18 & 100 & 48,852.69 & 3600.0 & 121.20 & 0.001 & 106.83 & 0.011 & 115.65 & 0.002 \\
        SD-5 & 6 & 32 & 24 & 100 & 111,061.52 & 3600.0 & 125.39 & 0.002 & 100.60 & 0.015 & 108.88 & 0.003 \\
        SD-6 & 6 & 32 & 24 & 100 & 58,009.11 & 3600.0 & 117.48 & 0.002 & \textbf{94.12} & 0.013 & 112.81 & 0.003 \\
        SD-7 & 6 & 40 & 30 & 100 & 329,125.32 & 3600.0 & 105.85 & 0.004 & \textbf{96.37} & 0.003 & 98.59 & 0.004 \\
        SD-8 & 6 & 48 & 36 & 100 & 468,476.71 & 3600.0 & 104.67 & 0.002 & 103.91 & 0.004 & \textbf{98.83} & 0.005 \\
        SD-9 & 6 & 48 & 36 & 100 & 166,905.92 & 3600.0 & 117.94 & 0.003 & \textbf{98.87} & 0.013 & 107.07 & 0.005 \\
        SD10 & 6 & 64 & 48 & 100 & 214,460.32 & 3600.0 & 114.24 & 0.015 & 99.99 & 0.010 & \textbf{99.86} & 0.008 \\
        SD11 & 6 & 80 & 60 & 100 & 1,298,915.34 & 3600.0 & 99.46 & 0.013 & \textbf{94.78} & 0.006 & 95.35 & 0.013 \\
        SD12 & 6 & 80 & 60 & 100 & 647,071.78 & 3600.0 & 104.97 & 0.022 & \textbf{96.73} & 0.029 & 100.69 & 0.010 \\
        SD13 & 6 & 96 & 72 & 100 & 921,796.96 & 3600.0 & 104.91 & 0.032 & 99.06 & 0.032 & \textbf{96.40} & 0.014 \\
        SD14 & 6 & 120 & 90 & 100 & 964,805.41 & 3600.0 & 103.98 & 0.052 & \textbf{97.83} & 0.039 & 101.50 & 0.010 \\
        SD15 & 6 & 144 & 108 & 100 & 1,367,958.17 & 3600.0 & 103.74 & 0.079 & 98.48 & 0.034 & \textbf{97.59} & 0.027 \\
        SD16 & 6 & 144 & 108 & 100 & 196,387.91 & 3600.0 & 111.23 & 0.045 & \textbf{95.95} & 0.009 & 100.59 & 0.032 \\
        SD17 & 6 & 160 & 120 & 100 & 2,492,351.39 & 3600.0 & 101.80 & 0.101 & \textbf{98.26} & 0.044 & 98.94 & 0.032 \\
        SD18 & 6 & 160 & 120 & 100 & 1,268,299.94 & 3600.0 & 105.08 & 0.100 & 99.18 & 0.015 & \textbf{97.24} & 0.033 \\
        SD19 & 6 & 192 & 144 & 100 & 1,815,221.39 & 3600.0 & 103.02 & 0.157 & 98.42 & 0.033 & \textbf{97.17} & 0.033 \\
        SD20 & 6 & 240 & 180 & 100 & 3,723,202.03 & 3600.0 & 102.21 & 0.293 & \textbf{98.66} & 0.050 & 99.71 & 0.060 \\
        SD21 & 6 & 288 & 312 & 100 & 793,041.57 & 3600.0 & 106.12 & 0.476 & \textbf{96.72} & 0.020 & 97.90 & 0.086 \\
        \midrule
        Average & 6 & 107 & 86 & 100 & 856,302.46 & 3600.0 & 108.86 & 0.067 & 100.43 & 0.019 & 101.99 & 171.432 \\
        \bottomrule
    \end{tabular}
    }
    \label{tab:sd_set_k6}
\end{table*}

\begin{table*}[htbp]
    \centering
    \caption{Comparison of \textsc{Appx} with \textsc{Greedy}, \textsc{Alg.1+}, and \textsc{Alg.2} on G Set}
    \resizebox{0.85\linewidth}{!}{ 
    \begin{tabular}{lcccccccccccc}
        \toprule
         & & & & & \multicolumn{2}{c}{\textsc{Appx}} & \multicolumn{2}{c}{\textsc{Greedy}} & \multicolumn{2}{c}{\textsc{Alg.1+}} & \multicolumn{2}{c}{\textsc{Alg.2}}\\
        \cmidrule(lr){6-7} \cmidrule(lr){8-9} \cmidrule(lr){10-11} \cmidrule(lr){12-13}
        Instance & \# Dep. & \# Cus. & \# Veh. &Cap. & OBJ & Time (s) & NS-value (\%) & Time (s) &  NS-value (\%) & Time (s) &  NS-value (\%) & Time (s)\\
        \midrule
        G-1 & 10 & 500 & 513 & 100 & 499,616.37 & 3600.0 & 116.88 & 7.844 & \textbf{99.32} & 845.608 & 101.79 & 3600.000 \\
        G-2 & 10 & 500 & 339 & 200 & 543,882.54 & 3600.0 & 104.15 & 2.163 & \textbf{99.46} & 123.574 & 102.57 & 3600.000 \\
        G-3 & 10 & 500 & 557 & 300 & 298,604.52 & 3600.0 & 105.92 & 5.511 & \textbf{98.39} & 164.443 & 103.97 & 3600.000 \\
        G-4 & 10 & 500 & 610 & 400 & 261,289.48 & 3600.0 & 100.99 & 4.874 & \textbf{98.32} & 109.207 & 105.45 & 3600.000 \\
        G-5 & 12 & 750 & 822 & 100 & 546,481.28 & 3600.0 & 116.79 & 19.198 & \textbf{98.90} & 3600.000 & 102.04 & 3600.000 \\
        G-6 & 12 & 750 & 757 & 200 & 435,573.00 & 3600.0 & 103.87 & 12.851 & \textbf{98.87} & 3600.000 & 103.99 & 3600.000 \\
        G-7 & 12 & 750 & 663 & 300 & 408,286.23 & 3600.0 & 103.93 & 10.715 & \textbf{97.95} & 3600.000 & 102.93 & 3600.000 \\
        G-8 & 12 & 750 & 746 & 400 & 343,026.94 & 3600.0 & 99.61 & 15.419 & \textbf{96.98} & 3600.000 & 102.76 & 3600.000 \\
        G-9 & 14 & 1000 & 798 & 100 & 639,220.60 & 3600.0 & 109.68 & 26.687 & \textbf{99.44} & 3600.000 & 102.46 & 3600.000 \\
        G10 & 14 & 1000 & 605 & 200 & 405,647.38 & 3600.0 & 104.88 & 19.131 & \textbf{98.39} & 3600.000 & 104.83 & 3600.000 \\
        G11 & 14 & 1000 & 857 & 300 & 502,720.97 & 3600.0 & 103.90 & 24.165 & \textbf{98.68} & 3600.000 & 103.55 & 3600.000 \\
        G12 & 14 & 1000 & 669 & 400 & 487,109.65 & 3600.0 & 107.31 & 17.697 & \textbf{97.59} & 3600.000 & 102.21 & 3600.000 \\
        G13 & 16 & 1250 & 1014 & 100 & 918,716.23 & 3600.0 & 109.93 & 40.998 & \textbf{98.63} & 3600.000 & 101.62 & 3600.000 \\
        G14 & 16 & 1250 & 619 & 200 & 539,266.94 & 3600.0 & 110.01 & 22.474 & \textbf{98.02} & 3600.000 & 103.36 & 3600.000 \\
        G15 & 16 & 1250 & 867 & 300 & 708,205.37 & 3600.0 & 109.90 & 32.585 & \textbf{97.79} & 3600.000 & 101.45 & 3600.000 \\
        G16 & 16 & 1250 & 952 & 400 & 581,455.42 & 3600.0 & 105.49 & 40.443 & \textbf{97.87} & 3600.000 & 102.06 & 3600.000 \\
        G17 & 18 & 1500 & 930 & 100 & 547,955.10 & 3600.0 & 103.27 & 50.590 & \textbf{97.68} & 3600.000 & 104.04 & 3600.000 \\
        G18 & 18 & 1500 & 975 & 200 & 774,259.43 & 3600.0 & 111.56 & 52.558 & \textbf{98.42} & 3600.000 & 102.46 & 3600.000 \\
        G19 & 18 & 1500 & 1030 & 300 & 706,122.34 & 3600.0 & 114.68 & 58.447 & \textbf{98.35} & 3600.000 & 102.96 & 3600.000 \\
        G20 & 18 & 1500 & 964 & 400 & 645,737.85 & 3600.0 & 112.60 & 51.371 & \textbf{97.65} & 3600.000 & 102.47 & 3600.000 \\
        \midrule
        Average & 14 & 1000 & 764 & 250 & 539,658.88 & 3600.0 & 107.74 & 25.786 & 98.31 & 2942.142 & 102.81 & 3600.000 \\
        \bottomrule
    \end{tabular}
    }
    \label{tab:g_set}
\end{table*}

\end{document}